\documentclass[11pt]{article}
\usepackage[text={17cm,22.1cm},centering]{geometry}
\usepackage{graphics}
\usepackage{amsmath}
\usepackage{comment}
\usepackage{xcolor}
\usepackage{float}
\usepackage{hyperref, color}
\usepackage[title]{appendix}
\def\ni{\noindent}
\def\ph{{\phantom{...}}}
\def\={\phantom{..} = \phantom{..}}

\def\+{\phantom{..} + \phantom{..}}
\def\>{\phantom{..} > \phantom{..}}
\def\<{\phantom{..} < \phantom{..}}
\def\-{\phantom{..} - \phantom{..}}

\def\srmo{{\sqrt{-1}}}
\def\oneq{{\frac{1}{4}}}

\def\bq{\begin{quote}}
\def\eq{\end{quote}}
\def\be{\begin{equation}}
\def\ee{\end{equation}}
\def\bar{\begin{eqnarray}}
\def\ear{\end{eqnarray}}
\def\no{\nonumber}

\def\half{{\frac{1}{2}}}

\def\Gartner{G{\"a}rtner}
\def\Sch{Schr{\"o}dinger}
\def\Schism{Schr{\"o}dingerism}
\def\Schist{Schr{\"o}dingerist}
\def\Schists{Schr{\"o}dingerists}

\def\Copist{Copenhagenist}
\def\Copists{Copenhagenists}

\def\wf{wavefunction}
\def\wfs{wavefunctions}

\def\rvs{random variables}

\def\MP{Measurement Problem}

\def\hpsi{\hat{\psi}}

\def\Plb{P\left[\,}
\def\Prb{\,\right]}
\def\sumS{{\sum_S\,|\psi|^2(S)}}
\def\sumi{{\sum_{i=1}^N}}
\def\ESl{{E_S(\lambda)}}
\def\hatZ{{\hat{Z}}}

\author{
	{\LARGE Leonardo De Carlo}\\
	\\
	\\
	Scuola Normale Superiore\\
	Piazza dei Cavalieri, 7, 56126 Pisa,\\
	Italia\\
	\texttt{leonardo\_d3\_carlo@protonmail.com}
	\and
	 {\LARGE W. David Wick}\\
	 \\
	 \\
	 Seattle,\\
	 USA\\
	\texttt{wdavid.wick@gmail.com}
}

\date{}
\title{\bf \huge On \Schist\ Quantum Thermodynamics\\[2in]}

\begin{document}
\maketitle

\pagebreak

\begin{abstract}
	
From the point of view of \Schism, a wavefunction-only philosophy, thermodynamics must be recast
in terms of an ensemble of wavefunctions, rather than classical particle configurations
or  ``found" values of Copenaghen Quantum Mechanics. Recapitulating the historical sequence, we consider here several 
models of magnets that classically can exhibit 
a phase transition to a low-temperature magnetized state.
We formulate wavefunction analogues including 
a ``\Schist\ QUantum Ising Model" (SQUIM), a ``\Schist\ Curie-Weiss Model" (SCWM), and others. 
We show that the SQUIM with free boundary conditions and distinguishable ``spins"
has no finite-temperature  phase transition, which we attribute to
entropy swamping energy. 
The SCWM likewise, even assuming exchange symmetry in the wavefunction (in this case the analytical argument is not totally satisfactory and we helped ourself with a computer analysis).  But a variant model with
``\emph{Wavefunction Energy}" 
(introduced in
prior communications about \Schism\ and the \emph{Measurement Problem}) does have a phase transition
to a magnetised state. The three results together suggest that magnetization  in large wavefunction spin
chains appears if and only if we consider indistinguishable particles and block macroscopic dispersion (i.e. macroscopic superpositions) by energy conservation.
Our principle technique involves transforming the problem to one
in probability theory, then applying results from Large Deviations, particularly the
\Gartner-Ellis Theorem. 
Finally, we discuss Gibbs {\em vs}. Boltzmann/Einstein entropy
in the choice of the quantum thermodynamic ensemble, as well as open problems.

\vspace{0.2cm}
\noindent{\em PhySH}:\, quantum theory, quantum statistical mechanics, large deviation \& rare event statistics.

\end{abstract}

\tableofcontents

\section{Background:The CIM and the CQUIM}

The classical Ising model (CIM)
features $N$ ``spins" or variables, which we write $S_1,...,S_N$, taking discrete, binary values, 
which for traditional purposes we will take to be $\pm\half$. 
To introduce dimensionality, we adopt a lattice of locations in d-dimensional space, 
say specified by d-tuples of integers in a finite box, $N$ sites in total.
 Next we associate
an energy with each configuration:

\be
E(S_1,...,S_N) = -2\, \sum_{i,j;\, n.n.}^N\,S_i\,S_j,
\ee

\ni where ``$n.n.$" stands for nearest neighbors on the lattice. (Including the factor of 2
simplifies some formulas we will need below.)
For example, with d=1 we have

\be
E(S_1,...,S_N) = -2\, \sum_{i=2}^N\,S_{i-1}\,S_i.
\ee

The minus sign is introduced so that the minimum energy appears at the ``all up" and the
``all down" states; introducing ``spin flips" (discordant n.n pairs) tends to increase the energy. 

We can now construct a Gibbs canonical ensemble with temperature $T$ by defining the average of some
functional of the spins to be 

\be
\left[\,F\,\right] \= Z^{-1}\,\sum_{S_1 = \pm\,1/2...,S_N = \pm\,1/2}\,\exp\left\{
- E(S_1,...,S_N)/kT\,\right\}\,F(S_1,...,S_N),
\ee

\ni where $Z$ is the normalizing constant

\be
Z \= \sum_{S_1 = \pm\,1/2...,S_N = \pm\,1/2}\,\exp\left\{
- E(S_1,...,S_N)/kT\,\right\}.
\ee

\ni (We have used $\left[...\right]$ to denote thermal average, 
as we can't use the usual angle brackets: $<...>$, because they will appear 
with a different meaning when we introduce \wfs.)
Equivalently, the sum over the $2^N$ states can be replaced by integration with respect to
the product of the binary measure assigning $\half$ to each of the cases $\pm\half$.  

Ising was interested in whether this model could represent a ferromagnet in the thermodynamic limit
($N \to \infty$) and hence
spontaneously magnetize at sufficently low temperature. 
Define the magnetic field generated by the N spins as:

\be
M \= \sum_{i=1}^N\,S_i.
\ee

From the up-down symmetry we deduce that $\left[\,M\,\right] = 0$,
so the average isn't the relevant question. 
There are two ways to continue. We can introduce + boundary
conditions, say by fixing $S_1 = S_N = +\half$, breaking the symmetry, and inquire as to whether,
below a critical temperature, 

\be
\lim_{N \to \infty}\,\left[\,M\,\right] > 0.
\ee

Another approach is to maintain the ``free" boundary conditions
and inquire into the behavior of the field-per-spin, $M/N$; does it exhibit
Central Limit Theorem (CLT) behavior; that is, $\approx$ O$(1/\sqrt{N})$? Which 
implies $\left[\,|M|\,\right] \approx \sqrt{N}$. If we interpret ``O$(N$)'' as macroscopic, 
we conclude that spontaneous magnetization didn't appear. The CLT would hold
if the correlations decayed exponentially:

\be
\left[\,S_i\,S_j\,\right] \approx \exp\{\,- \xi\,d(i,j)\,\};
\ee

\ni (for some constant $\xi > 0$, where $d(i,j)$ denotes distance between
 the lattice points labelled $i$ and $j$) 
but not if correlations decayed more slowly or even were bounded away from zero. In that
case, assuming below a critical temperature the dispersion 

\be
\left[\,\left(\,M/N\,\right)^2\,\right] \approx \hbox{constant} > 0,
\ee

\ni we would conclude that the limiting ensemble was a mixture of magnetized states 
and a phase transition had occurred.

Ising proved in his thesis of 1925 that the eponymous model did not magnetize in one dimension.
Subsequently, Peierls proved in 1936
that the model with $d=2$ does have a finite-temperature,
phase transition at some temperature $T_c$ to a magnetized state; then Onsager
solved the model exactly, including displaying formulas for $\left[\,|M|\,\right]$ and $T_c$.
The three-dimensional model has never been solved, but general arguments indicate that
such a phase transition appears for all models with $d\geq 2$.

By the \Copist\ QUantum Ising Model (CQUIM) we mean a special case of the Heisenberg model
with Hamiltonian operator expressed in terms of Pauli matrices as:

\be
 H \= -\, 2\,\sum_{i,j;n.n.}^N\,\sigma_{z;i}\,\sigma_{z;j}.
\ee

\ni Thus only the z-axis spins interact. 

In Copenhagen Quantum Mechanics one first diagonalizes the Hamiltonian
to discover the eigenvalues, call them $\{E_n\}$; then 
if the energy is ever measured, you will ``find one of the $E_n$, 
and the system jumps into the corresponding eigenstate". 
The realist assumption that a system actually exists exists 
in one of the eigenstates leads to contradictions
whenever certain pairs of observables are involved, 
such as position and momentum, or energy and time.
Hence the reliance on
textual formulations involving ``finding" rather than ``being". (John Bell made these points
eloquently, especially in his last paper ``Against `measurement'" \cite{Bell}.)   
    
Thus it would appear  that a thermal ensemble is constructed 
using these energy-measurement ``outcomes" of form:

\be
\left[\,F\,\right] \= Z^{-1}\,\sum_n\,\exp\{\,- E_n/kT\,\}\,<\psi_n|F|\psi_n>,
\ee

\ni where $\psi_n$ is the eigenstate corresponding to eigenvalue $E_n$ in the Hilbert space.
If we interpret the classical spins $S_i$ as pointing up or down the z-axis and
as labels of Hilbert space vectors, e.g., in Dirac notation, $|S_1,...,S_N>$,
they already define an eigenbasis of $H$ and the CQUIM reverts to the CIM. 
(If $H$ contains other operators not commuting with the $\sigma_z$, 
e.g. an external magnetic field in the x-direction, 
this is not the case and different basis must be found. 

Next we restrict to the d=1 case and reprise the computation of the dispersion 
of $M$ with ``free" boundary conditions, which we 
endeavored to imitate for a quantum model (next section). We focus on 
computing the probability generating function (pgf):

\be
Z_N(\beta;\lambda_1,...,\lambda_N) \= \sum_{S_1,...,S_N = \pm 1/2}\,
\exp\{\, - \,\beta\,E(S_1,...,S_N) + \lambda\cdot S\,\},
\ee

\ni where from now on $\beta = 1/kT$ and $\lambda\cdot S = \sum \lambda_i\,S_i$.
With a formula for $Z(\beta;\lambda)$ in hand we can then compute the dispersion by setting
all the $\lambda_i \equiv \lambda$ and taking two derivatives of $\log(Z)$ (let a prime denote
$d/d\lambda|_0$):

\bar
\left(\,\log(Z)\,\right)' \= \left[\,M\,\right];\,\,
\left(\,\log(Z)\,\right)'' \= \left[\,M^2\,\right] \- \left[\,M\,\right]^2.
\ear

The idea is to ``integrate out" the $N^{\hbox{th}}$ spin to get one step in a recursion.
Noting

\be
E(S_1,...,S_N) = -2\,S_{N-1}\,S_N -2\, \sum_{i=2}^{N-1}\,S_{i-1}\,S_i,
\ee

\ni we find

\bar
\no Z_N(\lambda_1,...,\lambda_N) = \half\,\exp\{\,\lambda_N/2\,\}\,Z_{N-1}(\lambda_1,
...,\lambda_{N-1} + \beta) \+  \half\,\exp\{\,- \lambda_N/2\,\}\,Z_{N-1}(\lambda_1,
...,\lambda_{N-1} - \beta).\label{receq}
\ear

To detect the pattern let's make another step:

\bar
\no Z_N &\=& \half\,\exp\{\,\lambda_N/2\,\}\,\left\{\,\half\,
\exp\{\,(\lambda_{N-1} + \beta)/2\,\}\, Z_{N-2}(\lambda_1,
...,\lambda_{N-2} + \beta)\right. \+\\
\no && \left. \half\,
\exp\{\,- (\lambda_{N-1} + \beta)/2\,\}\, Z_{N-2}(\lambda_1,
...,\lambda_{N-2} - \beta)\right\} \+\\
\no && \half\,\exp\{\,- \lambda_N/2\,\}\,\left\{\,\half\,
\exp\{\,(\lambda_{N-1} - \beta)/2\,\}\, Z_{N-2}(\lambda_1,
...,\lambda_{N-2} + \beta)\right. \+\\
\no && \left. \half\,
\exp\{\,- (\lambda_{N-1} - \beta)/2\,\}\, Z_{N-2}(\lambda_1,
...,\lambda_{N-2} - \beta)\right\}.\\
\ear

We can rewrite the last result in matrix form as:

\be
 Z_N(\lambda_1,...,\lambda_N) \= Y_N^{\dag}\,A_{N-1}\,\hat{Z}_{N-2},
\ee

\ni where

\be
Y_N \= \half\,\begin{pmatrix}
   \exp\{\,\lambda_N/2\,\} \\ 
   \exp\{\,- \lambda_N/2\,\} \\ 
   \end{pmatrix},\,\,
\hat{Z}_{N-2} \= \begin{pmatrix}
	Z_{N-2}(\lambda_1,...,\lambda_{N-2} + \beta) \\
	Z_{N-2}(\lambda_1,...,\lambda_{N-2} - \beta) \\
\end{pmatrix}
\ee

\be
A_{N-1} \= \begin{pmatrix}
           \exp\{\,(\lambda_{N-1} + \beta)/2\,\}  
           & \exp\{\,- (\lambda_{N-1} + \beta)/2\,\}\\  
           \exp\{\,(\lambda_{N-1} - \beta)/2\,\}  
           & \exp\{\,- (\lambda_{N-1} - \beta)/2\,\}\\ 
          \end{pmatrix}
\ee

Continuing, we can write 

\be
 Z_N(\lambda_1,...,\lambda_N) \= Y_N^{\dag}\,A_{N-1}\,A_{N-2}\,A_{N-3} ... A_{1}\,\hat{Z}_{1}.
\ee

We next set $\lambda_i \equiv \lambda$ to obtain:

\be
 Z_N(\lambda) \= Y^{\dag}\,A^{N-1}\,\hat{Z}_{1}.
\ee

Thus we are reduced to computing the $N^{\hbox{th}}$ power of a matrix.
$A$ is not symmetric but is nevertheless diagonalizable:

\be
A \= Q^{-1}\,D\,Q,
\ee

We leave the computation of $Q$ and $D$ to the reader (or look it up in any textbook on
classical thermodynamics). The upshot is:

\be
D \= \begin{pmatrix}
     \xi_{+} & 0 \\
      0 & \xi_{-} \\
    \end{pmatrix}
\ee

\ni and for $Z_N$:

\be
 Z_N \= Y^{\dag}\,Q^{-1}\,D^{N-1}\,Q\,\hat{Z}_{1}.
\ee
 
Writing out this result gives an expression of form:

\be
Z_N \= c^N\,\left\{\,f\,\xi_{+}^{N-1} \+ g\,\xi_{-}^{N-1}\,\right\},\label{Zresult}
\ee

\ni where $c,f,g,\xi_{-},\xi_{+}$ all depend on $\lambda$.  
Finally, we have to examine:

\be
\frac{1}{N^2}\,\log(Z_N)'' \= \frac{1}{N^2}\,\left\{\,\frac{Z_N''}{Z_N} \- \frac{(Z_N')^2}{Z_N^2}
\,\right\}.\label{difeq}
\ee

The largest factor in (\ref{Zresult}) is $\xi_{+}^{N-1}$, which tends to infinity. In terms of
(\ref{difeq}) in which this factor is differentiated, 
including $(N-1)\,\xi_{+}^{N-2}\,\xi_{+}'$ and
$(N-1)\,(N-2)\,\xi_{+}^{N-3}\,\xi_{+}'^2$,  
the term has lower order and the denominator
wins the race. In terms where it is not, the ratio tends 
to a constant and the prefactor sends it to zero. Hence the dispersion of the mean tends to zero
and the CIM with $d=1$ does not exhibit spontaneous magnetization.

\subsection{Mathematical results}

The objective of this work is to understand if it is possible to construct 
physically meaningful wave-mechanical magnetic models and to explore the implications. In section \ref{sec:SQUIM} we consider distinguishable particles, 
namely we take the full $ 2.2^N $-sphere as state space 
without imposing any exchange symmetry on the wavefunctions.
In this case the models 
will not magnetize (except at $ 0 $ temperature): $ \lim_{N\to \infty}\,[\,\{\, |m(\psi)| > 
\epsilon\,\}\,]_{\beta} = 0 $ for any $ \epsilon $, 
where  $ m(\psi) = <\psi|\,\sum\,S_i/N\,|\psi> = 
\sum\,|\psi(S)|^2(\,\sum\,S_i\,)/N $ and $ [\cdot]_\beta $ 
is the ensemble thermal average. 

In section \ref{sec:ModwithWFE} we introduce a  
generalized  Hamiltonian framework, 
as developed in \cite{MP1}, assuming a  non-linearity 
that penalizes large superpositions in large objects. This non-linearity makes energetically impossible configurations 
corresponding to macroscopic cat states\footnote{For a modern mathematical 
	definition of cat states see \cite{MP1}.}. 
Even in this case, with distinguishable particles 
no phase transition appears (except at zero temperature).

In section \ref{SCWsect} and \ref{sec:SCWwithWFE} 
we specialize to the mean field case with exchange symmetry.
With this choice, the dimensionality of  the Hilbert state space is greatly reduced and 
the  model manifests magnetization at finite temperatures, when Wavefunction Energy is included.
(In this case but without WFE, showing that there is no phase transition 
assumed some properties 
of a certain two-argument function for which we provide computer-generated evidence.) 
To our knowledge, this is the simplest wavefunction model in which a phase transition appears. 

In section \ref{leonardo}, we discuss the choice of ensembles 
with uniform measure, as opposed to other possibilities. 
In particular we introduce a base measure that includes 
a quantum Boltzmann-Einstein entropy and discuss how the 
usual ensembles should be recovered in a suitable limit.  

\vspace{0.6cm}

The paper is organized as follows: in each section the primary results are described
but the mathematical derivations are relegated to appendices.
Our principle technique involves transforming the problem to one
in probability theory, then applying results from Large Deviations, particularly the
\Gartner-Ellis Theorem.

\subsection{Abbreviations}
The following abbreviations are used in this manuscript:\\

\noindent 
\begin{tabular}{@{}ll}
	
	CCW & classical Curie-Weiss\\
	CIM & classical Ising model \\
	COM & center-of-mass\\
	CQUIM & \Copist\ QUantum Ising Model \\
	ESM & Exactly Solvable Model\\
	SCW & \Schist Curie-Weiss\\
	SQUIM &\Schist\ QUantum Ising Model\\
	UIF & Useful Integral Formula\\
	WFE & Wavefunction Energy\\
\end{tabular}

\section{The SQUIM}\label{sec:SQUIM}

As for the CIM, we consider $N$ spins $S_1,...,S_N$ taking values $\pm\half$. 
But the state is not now a configuration but a \wf\ on configurations, 
i.e., of form $\psi(S_1,...,S_N)$ where each value is a complex number. 
Thus $\psi$ lies in a space of $2.2^N$ real dimensions. 
It is unnatural to \Schists\ to insist on ``finding" 
an energy eigenvalue; for us, the energy of a \wf\ $\psi$
 is what in Copenaghen Quantum Mechanics is called ``the expected energy".
In particular, the energy $E(\psi)$ contains
contributions from all spin configurations (``superpositions"). (As quantum mechanists remind us:
``Remember, all the states are there.") 
The Gibbs canonical ensemble is not based on a product measure. 

The pgf we wish to evaluate now takes the form:
\be
Z_N(\beta;\lambda_1,...,\lambda_N) \= \int_{||\psi_N|| = 1}\,d\psi_N\,\exp
\left\{\,- \beta\,E_N(\psi_N) + \Lambda_N(\psi_N)
\,\right\}.\label{Zdef}
\ee
\ni ($||\psi||^2 = \sum\,|\psi(S_1,...,S_N|^2.$) 
The integral is over the unit sphere in the above-stated number of dimensions. 

The motivation for the base integral over normalized \wfs\ 
differs from the \Copist, in which $\psi$ represents a probability
distribution on spin configurations. Rather, \Schist\ reasoning is that we
do not want to compare states on the basis of normalizations, e.g., 
one with norm 0.1 and another with norm 100, but solely by their respective energies. 
Moreover, both linear Quantum Mechanics and the nonlinear generalization one of us discussed
in connection with the Measurement Problem, \cite{MP1}, preserve the norm, and the latter
raises the possibility that the flow is ergodic or chaotic, 
with a possible interest in a classical justification for the Second Law of Thermodynamics. 
Thus we would not wish to drop the normalization. 

Limiting ourselves to $d = 1$ as in the previous section: 
\bar
\no E_N(\psi_N) \= - <\psi_N|\,\sum_{i=2}^N\,S_{i-1}\,S_i\,|\psi_N>=- \sum\,|\psi_N(S_1,...,S_N)|^2\,
\sum_{i=2}^N\,S_{i-1}\,S_i,
\ear
\ni and
\bar
\no \Lambda_N(\psi_N;\lambda_1,...,\Lambda_N) \= <\psi_N|\sum_{i=1}^N\,\lambda_i\,S_i\,|\psi_N>= \sum\,|\psi_N(S_1,...,S_N)|^2\sum_{i=1}^N\,\lambda_i\,S_i.
\ear

We began by trying to imitate Ising's 1925 and develop a recursion, as reprised
in the previous section.
Accordingly, we divided the \wf\ into two sets, corresponding to the last spin $+\half$
or $-\half$, leading to the splitting
\be 
\psi_N \= \left(\,\psi_{+,N-1}(S_1,...,S_{N-1}),
\psi_{-,N-1}(S_1,...,S_{N-1})\,\right).
\ee
\ni This splitting leads to the decomposition of the exponent in the Gibbs factor:
\bar
\no && - \beta\, E_N(\psi_N) + \Lambda_N(\psi_N) \= \\
\no && \sum\,|\psi_{+,N-1}(S_1,...,S_{N-1})|^2\,\left\{ \beta\,
\sum_{i=2}^{N-1}\,S_{i-1}\,S_i + \half\lambda_N + \sum_{i=1}^{N-2}\,\lambda_i\,S_i\,
+ [\,\lambda_{N-1} + \half\beta\,]\,S_{N-1}\right\}\\
 && \+ \left[\,\half\,\leftrightarrow\,- \half\,\right] \=\\
\no && \half\,\lambda_N\,||\psi_{+,N-1}||^2 + 
\beta\,E_{N-1}(\psi_{+,N-1}) + \Lambda_{N-1}(\psi_{+,N-1};\lambda_1,...,\lambda_{N-1} 
+ \half\beta)  \+ \left[\,\half\,\leftrightarrow\,- \half\,\right]. 
\ear
Since 
\be
||\psi_{+,N-1}||^2 \+ ||\psi_{-,N-1}||^2 \= 1,
\ee
\ni we can write: $\cos(\theta) = ||\psi_{+,N-1}||$ and 
$\sin(\theta) = ||\psi_{-,N-1}||$ for some $\theta$ with $0 \leq \theta \leq \pi/2$,
then rewrite the above decomposition in terms of normalized components, e.g.,
$ \psi_{+,N-1} = \cos(\theta)\hat{\psi}_{+,N-1}$, with 
$||\hat{\psi}_{+,N-1}||^2 = 1$. This yields a splitting of form
\be 
\psi_N \= \left(\,\cos(\theta)\,\hat{\psi}_{+,N-1}(S_1,...,S_{N-1}),
\sin(\theta)\,\hat{\psi}_{-,N-1}(S_1,...,S_{N-1})\,\right).
\ee
This decomposition is a special case of so-called ``polyspherical coordinates", \cite{russians}.
In particular, the invariant measure can be written\footnote{See formula (4), section 9.1.9, of
\cite{russians}. We learned about this system, which differs from the usual spherical coordinates,
on Wikipedia; on the page entitled "n-sphere", around 12/1/2020.}:
\be
d\psi_N \= d\hat{\psi}_{+,N-1}\,d\hat{\psi}_{-,N-1}\,c_N\,\cos^{k(N-1)}(\theta)
\,\sin^{k(N-1)}(\theta)\,d\theta,
\ee
\ni where $k(N-1) = 2.2^{N-1}$ and the coefficients $c_N$ we need not indicate, 
since at the end of the computation we will take logarithms 
and differentiate with respect to the $\lambda$'s, so 
they drop out.

Plugging all into (\ref{Zdef}) yields
\def\ctt{{\cos^2(\theta)}}
\def\stt{{\sin^2(\theta)}}
\bar
\no Z_N(\beta;\lambda_1,...,\lambda_N) &\=&c_N\,\int_0^{\pi/2}\,d\theta\,\cos^{k(N-1)}(\theta) 
\,\sin^{k(N-1)}(\theta)\,\times \exp\{\,\half\lambda_N\,(\,\ctt - \stt\,)\,\}\,\times \\
\no && Z_{N-1}(\beta\ctt;\ctt\,\lambda_1,...,
\ctt\,\lambda_{N-2},\ctt\,[\,\lambda_{N-1} + \half\beta\,]\,)\,\times\\
\no && Z_{N-1}(\beta\stt;\stt\,\lambda_1,...,
\stt\,\lambda_{N-2},\stt\,[\,\lambda_{N-1} - \half\beta\,]\,)\\
&&
\ear
\ni which is the sought-after recursion.
Unfortunately, we were unable to obtain a useful solution of this nonlinear recursion.
(The polyspherical coordinates will help in a simpler problem; see Math Appendix \ref{app:models no stat}.) 
Therefore, we pursued a probability-style proof. The trick is to replace the random point on the
sphere with 
\be
\psi(S) \longrightarrow \frac{\chi(S)}{\sqrt{\sum\,|\chi(S)|^2}},
\ee
\ni where $\{\,\chi(S): S = (S_1,...,S_N)\,\}$ are $2^N$ complex, or 2.$2^N$ real, numbers  
distributed as i.i.d. standard (mean zero, norm one) Gaussians. 
This procedure defines a measure on the sphere 
and the rotational symmetry of the standard Gaussian measure (covariance matrix: the identity)
then assures that we are considering the same distribution of wavefunctions. By this transformation
and various Lemmas reducing the problem to finding probabilities of rare events (called
``Large Deviations" theory in the trade) we were able to prove the following theorem.
Let
\bar
\no && m(\psi_N) = <\psi_N|\,\sum\,S_i/N\,|\psi_N> = \sum\,|\psi_N(S)|^2(\,\sum\,S_i/N\,);\\
\no && [\,F(\psi_N)\,]_{\beta}  
\= \int_{||\psi_N|| = 1}\,\,d\psi_N\,\exp\{ - \beta\,E_N(\psi_N)\,\}\,F(\psi_N)/Z_N.\\
&&
\ear

\begin{quote}
\centerline{\bf Theorem One}\label{th:1}

The SQUIM in any dimension has zero magnetization for any finite temperature (except for  $ T=0 $) in the thermodynamic
limit:
\be
\lim_{N\to \infty}\,[\,\{\, |m(\psi)| > \epsilon\,\}\,]_{\beta} = 0,
\ee
\ni for any $\epsilon > 0$.
\end{quote}

For the proof, see the Math Appendix \ref{app:models no stat}. 
(One can deduce from the arguments in this Appendix that 
neither a mean field interaction  will change the situation.)
 
This model does exhibit zero-temperature
magnetization. As $\beta \to \infty$, the probability distribution becomes concentrated on
states of minimal energy. These have the form:
\be
\psi(\theta,\alpha) = \cos(\theta)\,\psi_{+} \+ \sin(\theta)\,e^{\srmo\,\alpha}\,\psi_{-},
\ee
\ni where $\psi_{\pm}$ denotes a wavefunction concentrated on all up, respectively all down,
spins. Hence at $T = 0$ the magnetization takes the form:
\bar
 \int_0^{\pi/2}\,d\theta\,f(\theta)\,<\psi(\theta,\alpha)|\frac{M}{N}|
\psi(\theta,\alpha)>^2 \= \frac{1}{4}\,\int_0^{\pi/2}\,d\theta\,f(\theta)\,\cos^2(2\theta) > 0.
\ear

\section{Models assuming Exchange Symmetry and the \Schist\ Curie-Weiss Model.\label{SCWsect}}

The model we called the ``\Schist\ QUantum Ising model" is peculiar 
from the perspective of the history of quantum theory, 
as we have omitted any assumption about invariance of \wfs\ under
exchange of arguments. 
Since the proof implies that  omitting this invariance we can not have phase transitions for any magnetic interaction, we include it starting from the first case where one expects such phenomena, namely the Curie-Weiss model.  

The classical Curie-Weiss (CCW) model assumes that the system creates an overall or mean field,
which interacts with every spin. The CCW energy is:
\be
E_{CW} \= -\frac{1}{N}\, \left(\,\sum_{i=1}^N\,S_i\,\right)^2.
\ee 

We are going to replace spin configurations by \wfs\ with exchange symmetry; in the ensemble we consider only symmetric wavefunctions, while one should consider also antisymmetric. We consider  $ \pm 1 $ as spin values, while other ones could be
considered (e.g   -1, 0, +1, \cite{Yamada,YYDX}).  We will sustain two, reasoning that the lower dimensionality
and mean field interaction defines the first case to study 
whether phase transitions appear or not. 
Models with higher dimensions, meaning higher entropy, or local\footnote{Both the concept and the mathematical definition 
of nearest neighbors for indistinguishable particles appears somewhat problematic in wavefunction
models.}  interactions presumably
are even less likely to exhibit such transitions.
In standard reference \cite{WL}, a qu-bit is considered 
a quantum two-level system, without the orbital component of its wavefunction,  and  many qu-bits having  the exchange symmetry property.( In \cite{53qubits} they  consider interacting qu-bits. IBM \cite{IBM} says that titanium atoms can represent  qu-bits and they project to make a complex of many of them to observe how the collective behavior changes,  this might be a situation described by the SCW model.)

\def\Ael{A_{\hbox{ellipsoid}}}
\def\Asp{A_{\hbox{sphere}}}

Note that with two levels the CW energy depends only on the number, call it `$n$', of ``up" spins.
Thus we are lead to introducing \wfs\ that depend only on `$n$'.
Before introducing such \wfs\, here is another (besides polyspherical coordinate systems)
way to define surface integrals over spheres or ellipsoids, that physicists prefer:
On N-dimensional Euclidean space,
define, given some $a_i > 0$:
\bar
 &&\int_{\{\,\sum x_i^2\,a_i = 1\,\}}\,\prod_{i=1}^N\,dx_i F(x_1,...,x_N) \= \\
\no && \lim_{\sigma \to 0}\,\Ael^{-1}\,\int\,\prod\,dx_i\,\exp\{\,- (\,\sum\,x_i^2\,a_i - 1\,)^2/\sigma^2\,\}
\,F(x_1,...,x_N)/W(\sigma),\label{intdef}
\ear
\ni where ``$\Ael$" stands for the surface area of the ellipse $\{\,\sum\,x_i^2\,a_i = 1\,\}$ and
\be
W(\sigma) \=\int\,\prod\,dx_i\,\exp\{\,- (\,\sum\,x_i^2\,a_i - 1\,)^2/\sigma^2\,\}.
\ee
\ni Thus we utilize an approximate Dirac delta-function 
becoming concentrated in the limit on the ellipsoid.

Now let us make a change of variables by defining: $y_i = x_i/
\sqrt{a_i}$ in \eqref{intdef}. A Jacobian derivative of $\prod \sqrt{a_i}$
appears in numerator and denominator and so cancels, yielding the theorem:

\begin{quote}
{\bf Lemma} Let $\Asp$ be the ``surface area" of the N-sphere. Then:
\bar
 \int_{\{\,\sum x_i^2\,a_i = 1\,\}}\,\prod_{i=1}^N\,dx_i F(x_1,...,x_N) \= \left(\,\frac{\Asp}{\Ael}\,\right)\,\int_{\{\,\sum y_i^2 = 1\,\}}\,
\prod_{i=1}^N\,dy_i F(\sqrt{a_1}\,y_i,...,
\sqrt{a_N}\,y_N).\label{theorem}
\ear
\end{quote}

We want to define an ensemble and corresponding integrals for symmetric
\wfs. Each such \wf\ is uniquely given by its common value, call it $\hpsi(n)$, on
the set of configurations with $n$ ``up" spins. We have then:
\bar
\no ||\psi||^2 &\=& \sum_{n=0}^N\,C(N,n)\,|\hpsi(n)|^2=1\\
\no E(\psi) &\=& \sum_S\,|\psi(S)|^2\,E_S= \sum_n\,\sum_{\{\#S = n\}}\,|\psi(S)|^2\,E_S\\
\no &\=& \sum_n\,C(N,n)\,|\hpsi(n)|^2\,\sum_{\{\#S = n\}}\,E_S/C(N,n)= \sum_n\,C(N,n)\,|\hpsi(n)|^2\,E_n.\\
&&
\ear
\ni Here ``$S$" stands for a spin configuration, ``$\#S$" for how many ``up" spins it contains,
 and $C(N,n)$ 
is the number of combinations with  $n$ spins up.
Next we let
\be
\phi(n) \= \sqrt{C(N,n)}\,\hpsi(n),
\ee
\ni and apply the theorem from above section, yielding:
\bar
\no && \int_{\{\,\sum\,|\hpsi(n)|^2 = 1\,\}}\,
\exp\left\{- \,\sum\,C(N,n)\,|\hpsi(n)|^2\,E_n\,\right\} \=\\ 
 && \frac{\Asp}{\Ael}\,\int_{\{\,\sum\,|\phi(n)|^2 = 1\,\}}
\,\exp\left\{- \,\sum \,|\phi(n)|^2\,E_n\,\right\}.\label{theoremapp}
\ear
\ni Note that the prefactor in (\ref{theoremapp}) is not in the integrand, not a function of
$\phi$, just a constant depending on $N$. 
Thus it plays no role in computing, e.g., the magnetization. 

With these prepatory remarks we can define our symmetric-\wf\, SCW model.
The magnetic energy associated to $\phi$ becomes:
\be
E_{CW}(\phi) \= - \frac{1}{N}\, \sum_{n=0}^N\,|\phi_n|^2\,\left(\,N-2n\,\right)^2.
\ee
Our SCW model is then defined by:
\bar
 [\,F(\phi)\,]_{\beta} = \int_{\{||\phi||^2 = 1\}}\,d\phi\,\exp\{ -\beta\,E_{CW}(\phi) \}\,
F(\phi)/Z; Z = \int_{\{||\phi||^2 = 1\}}\,d\phi\,\exp\{ -\beta\,E_{CW}(\phi) \}.
\ear
The choice of the uniform base ensemble might be questioned; we leave the issue to
section \ref{leonardo}.

We state this theorem:

\begin{quote}
\centerline{\bf (Numerical) Theorem Two}
For the SCWM at any finite temperature (except for $ T=0 $) and for any $\epsilon > 0$:
\be
 \lim_{N\to\infty}\,[\,\{\,|m(\phi)| > \epsilon\,\}\,]_{\beta} = 0.
\ee
\end{quote}
Thus the SCWM
with the standard energy expression has no finite-temperature phase transition.

\vspace{0.5cm}
Using Large Deviations techniques from probability theory, we reduced the proof
of Theorem Two to a study of one real function of two variables, for which we supply exact formulas
(See Math Appendix \ref{app:SCWnoWFE}). 
Some of the properties of the rate functionals, needed for the proof, 
are displayed via computer using a  statistical sampling from the domain of this function within our computational abilities.
Thus we can claim only a computer-assisted proof.   A complete analytical solution requests to solve an involved optmization problem, which would need another paper to develop some taylored analytical and numerical techniques for the problem and we supply a scheme useful to reach a complete proof.

\section{Models with Wavefunction Energy} \label{sec:ModwithWFE}

In previous papers the second auther explored nonlinear \Schist\ quantum mechanics 
as a potential solution to the \MP  \cite{MP1,MP2}. The proposal involved adding 
nonquadratic terms to the Hamiltonian, i.e., to the energy. We will refer to these 
terms as representing ``wavefunction energy" (WFE). 
These terms were proposed to affect only macroscopic objects 
(and otherwise be too small to matter). 
The WFE in the present work will be a functional of the wavefunction addressing 
dispersion of the total spin, instead of center-of-mass or total momentum (as in \cite{MP1}). 
It is not clear if this indicates a more general mechanism 
than the one of \cite{MP1,compatibility}, or just reflects 
the lack of spatial coordinates in our models of magnetism. 
We return to consider these issues in the Discussion section.
Our principles to modify the quantum Hamiltonian are:
\begin{itemize}
	\item[a)] The modification has to be negligible at microscopic level but it becomes large at macroscopic level to block macroscopic dispersion because of "configurational cost". For the present case macroscopic dispersion of the spins (better explained later);
	\item[b)] The norm of $ \psi $ and the energy of a closed system have to be conserved (the same for the momentums);
	\item[c)] No extra terms are added to the evolution of the center of mass $ X(\psi)= <  \psi |   \frac{1}{N}\underset{j=1}{\overset{N}{\sum}}x_j |\psi >   $  of a closed system.
\end{itemize}
These characteristic are satisfied when the  evolution of the quantum state is
\begin{equation*}\label{eq:nonlin}
i\hbar\frac{\partial \psi}{\partial t}=\frac{\partial}{\partial\psi^* }E(\psi).
\end{equation*}
with
\be
E(\psi) \= E_{QM}(\psi) \+ E_{WFE}(\psi),
\ee
\begin{equation}\label{eq:genwfe}
E_{WFE}(\psi)= wN^2 D(\psi),\,\, D(\psi):=  <\psi|\,\big(\,\overset{N}{\underset{i=1}{\sum}}\,O_i - <\psi|\,\overset{N}{\underset{i=1}{\sum}}\,O_i\,|\psi>\,\big)^2\,|\psi>
\end{equation}
where $ w $ is a very small positive constant   and the $ O_i$'s are  self-adjoint operators such that $\left(\overset{N}{\underset{i=1}{\sum}}\,O_i  \right)^2$ is self-adjoint too, see the proofs in \cite{MP1}.
For our statistical ensembles  the first term $ E_{QM}(\psi) $ now incorporates the usual spin-spin and spin-external field interactions, 
while the second $ E_{WFE}(\psi) $ takes the form:
\bar\label{eq:WFEspin}
\no E_{WFE}(\psi) &\=& w N^2\,D(\psi);\\
\no D(\psi) &\=& \frac{1}{N^2} <\psi|\,\big(\,\sum\,S_i - <\psi|\,\sum\,S_i\,|\psi>\,\big)^2\,|\psi>\\
\no &\=& \frac{1}{N^2}\left[<\psi|\,\big(\,\sum\,S_i\,\big)^2\,|\psi> - <\psi|\,\sum\,S_i\,|\psi>^2\right]\\
&&
\ear
Here  we have written the sum of spins rather than $M$ 
to emphasize the interpretation as the ``dispersion of the center-of-spin". The straightforward generalization of
\eqref{eq:WFEspin} to models with spatial coordinates is $ O_i = L_i + S_i $ in \eqref{eq:genwfe} and $ [w]= kg^{-1}m^{-2} $. The original proposal \cite{compatibility} is on the
momentum $ P_i $ . The general idea behind \eqref{eq:genwfe} derived in part 
from an interview with Hans Dehmelt, see chapter
15 in \cite{Faris-Wick}. 
He explained to one of the authors that classical physics applies when
for some reason wavefunctions cannot spread out (forming cats). 
For instance, classical-like orbits in electromagnetic traps are due
to the trapping fields. The  mechanism \eqref{eq:genwfe} proposes
a universal self-trapping forbidding macroscopic dispersion, so to speak.  (The ideal experimental test
to falsify WFE is described in \cite{MP3}.)

\vspace{0.8cm}
Let us first examine the $T=0$ case of the SQUIM, WFE version. Again, the case reduces to
support on \wfs\ of form:

\be
\psi(\theta,\alpha) = \cos(\theta)\,\psi_{+} \+ \sin(\theta)\,e^{\srmo\,\alpha}\,\psi_{-},
\ee

Plugging into the definition of $D$ yields:

\be
E_{WFE}(\psi) = \frac{N^2}{4}\,\sin^2(2\,\theta).
\ee

\ni Since by adding a trivial constant (which doesn't alter the measure) 
the usual energy can be rewritten:

\be
E_{QM}(S_1,...,S_N) = \sum_{i,j,n.n.}\,\left(\,S_i - S_j\,\right)^2,
\ee

\ni both forms of energy have the same sign (positive). So in the presence of the WFE terms
the support is reduced to the cases with $\theta = 0$ or $\theta = \pi/2$; hence the ensemble
becomes a mixture of two magnetized states, as for the CIM. ``Classical" behavior is restored.

Thus, the interesting question arises as to whether this conclusion 
will persist at nonzero temperatures. 
Here follows some ideas about the situation.

When $T > 0$, the suggestion is that the effect of $E_{WFE}(\psi)$ is to suppress \wfs\ with
$E_{WFE}(\psi) \approx$ O$(N^2)$---e.g., the above superposition---but not if O$(N$), 
like the usual (quadratic) terms.
However, $E_{WFE}(\psi) \approx$ O$(N)$ suggests decay of correlations and CLT behavior, 
which in the classical case limits macroscopic magnetization. On the other hand, if $d 
\geq 2$ the presence of $E_Q(\psi)$ pushes in the opposite direction!

However, the ``correlations" involved in the dispersion and the magnetization are different,
so we must be cautious. Writing $[\cdot]_C$ for classical thermal average over spins 
and $[\cdot]_{SQ}$ 
over \wfs:

\be
[\,F\,]_C = Z_{N,C}^{-1}\,\sum_{S_1=\pm \half,...,S_N = \pm\half}\,
\exp
\left\{\,- \beta\,E_S\,) 
\,\right\}\,F_S,
\ee

\be
[\,F\,]_{SQ} \= Z_{N,SQ}^{-1}\, \int_{||\psi_N|| = 1}\,d\psi_N\,\exp
\left\{\,- \beta\,E(\psi_N)\, 
\,\right\}\,F(\psi),
\ee

\ni we can write for the correlations:

\be
\hbox{CIM}: \ph\ph [\,S_i\,S_j\,]_C;
\ee

\be
\hbox{SQUIM}: \ph\ph [\,<\psi|\,S_i\,|\psi>\,<\psi|\,S_j\,|\psi>\,]_{SQ};
\ee

For the magnetizations:

\be
\hbox{CIM}: \ph\ph [\,\left(\,\sum\,S_i\,\right)^2\,]_C \=
 \sum_{i,j}\,[\,S_i\,S_j\,]_C;
\ee

\be
\hbox{SQUIM}: \ph\ph [\,\left(\,\sum\,<\psi|\,S_i\,|\psi>\right)^2\,]_{SQ} \=
 \sum_{i,j}\,[\,<\psi|\,S_i\,|\psi>\,<\psi|\,S_j\,|\psi>\,]_{SQ}.
\ee

By contrast, 

\bar
\no E_{WFE}(\psi) &\=& <\psi|\,\big(\,\sum\,S_i - <\psi|\,\sum\,S_i\,|\psi>\,\big)^2\,|\psi>\\
\no  &\=& \sum_{i,j}\,\left\{\,<\psi|\,S_i\,S_j\,|\psi> - <\psi|\,S_i\,|\psi>\,
<\psi|\,S_j\,|\psi>\,\right\},\\
&&\label{Deqn}
\ear

\ni which is rather different. Clearly, the ``correlations" suppressed by including $E_{WFE}(\psi)$
in the Gibbs factor concern single-wavefunctions and whether they imply a cat, while the
correlations important for magnetization are thermal.

Thus the situation may be delicate: the presence of the WFE terms may permit 
transition to a low-temperature magnetized state, 
perhaps altering the critical temperature $T_c$ 
(but we will see this is not the case). 

Before proceeding, let us ask: could these models be regarded as ``continuous spin" models,
in some sense, and therefore covered by the extensive literature on that topic?
First, consider \wfs\ that have $E_{WFE}(\psi)= $ O$(N)$. One class such are the product
states:

\def\fhalf{{1/2}}
\be
\hbox{P.S.} \= \left\{\,\psi:\ph \psi(S_1,...,S_N) = \prod_{i=1}^N\,\psi_i(S_i)\,\right\}.
\ee

\ni These states are normalized provided:

\be
|\psi_i(\fhalf)|^2 + |\psi_i(-\fhalf)|^2 \= 1,\ph \hbox{for all i = 1,..., N}.
\ee 

An easy computation gives for these states:

\bar
\no E_{WFE}(\psi) &\=& \oneq\,\sumi\,\left\{\,|\psi_i(\fhalf)|^2 + |\psi_i(-\fhalf)|^2\,\right\} \-
\oneq\,\sumi\,\left\{\,|\psi_i(\fhalf)|^2 - |\psi_i(-\fhalf)|^2\,\right\}^2\\
\no  &\=& \frac{N}{4}\, \-
\oneq\,\sumi\,\left\{\,|\psi_i(\fhalf)|^2 - |\psi_i(-\fhalf)|^2\,\right\}^2 \leq \frac{N}{4}.\\
&&
\ear

For the magnetization of these product states we find:

\bar
 <\psi|\,\sumi\,S_i\,|\psi> \=
\sumi\,\sum_{S_i = \pm 1/2} |\psi_i(S_i)|^2\,S_i\= \sumi\,V_i,
\ear

\ni where the ``spins" $V_i$ are defined here.

For the usual (quadratic) energy:

\be
- <\psi|\,\sum_{i,j n.n.}\,S_i\,S_j\,|\psi> \=  
- \sum_{i,j n.n.}\,V_i\,V_j.
\ee

As the $V_i$ take values in the interval $[-\half,\half]$, we detect a continuous-spin 
Ising-type model developing. However, it is not clear what the base measure on the $V_i$'s,
call it $P(V_1,...,V_N)$,
should be. Assuming it arises from the spherical-measure by concentration on the set P.S.
(which has measure zero, so as our SQUIM measure becomes singular), we see that it should be
exchange- and reflection-invariant (meaning under the overall spin-flip operation). However,
those conditions do not imply that 

\be
P(V_1,...,V_N) \= \prod_{i=1}^N\,P_i(V_i).
\ee

For a continuous-spin model most authors would adopt the product form with some symmetrical
marginal distribution, say uniform on $[-\half,\half]$. Theorems have been proven yielding
phase-transition for sufficently large $d \geq 2$. 

However, the inclusion of the WFE terms cannot be presumed to limit the ensemble
to product states (at finite temperature). This can be seen by expanding the first line in  
(\ref{Deqn}):
\bar
\no  && \sum_{i,j}\,<\psi|\,\big(\,\,S_i - <\psi|\,S_i\,|\psi>\,\big)\,
\big(\,\,S_j - <\psi|\,S_j\,|\psi>\,\big)\,|\psi> \= \\
\no  && \sum_{i}\,<\psi|\,\big(\,\,S_i - <\psi|\,S_i\,|\psi>\,\big)^2
\,|\psi> \+ \\
\no  && \sum_{i \neq j}\,<\psi|\,\big(\,\,S_i - <\psi|\,S_i\,|\psi>\,\big)
\big(\,\,S_j - <\psi|\,S_j\,|\psi>\,\big)\,|\psi>\\
&&
\ear
\ni Only the first term survives in a product state, for which $E_{WFE}(\psi) \approx$ O$(N)$.
But the latter can still hold for nonproduct states, e.g., ones for which the ``correlations"
in the last line above are nonpositive. Hence we cannot directly rely on the literature
about classical continuous-spin models.

We summarize how  $ D(\psi) $ works. It is of order $ 1 $ on \emph{cat states}, i.e.  close\footnote{The concept can be made precise introducing a suitable spherical distance.} to  \begin{equation}\label{eq:statigatto}
\psi=\frac{1}{\sqrt{2}}\psi_{+}+ \frac{e^{\sqrt{-1}\alpha}}{\sqrt{2}}\psi_{-},
\end{equation}
and therefore $ E_{WFE}(\psi) $ of order $ w N^2 $, of order $ 1/N $ on \emph{product  states}, i.e.  close to 
\def\fhalf{{1/2}}
{ \be
	\hbox{P.S.} \= \left\{\,\psi:\ph \psi(S_1,...,S_N) = \prod_{i=1}^N\,\psi_i(S_i)\,\right\},\,|\psi_i(\fhalf)|^2 + |\psi_i(-\fhalf)|^2 \= 1,\ph \forall\,\,i.
	\ee}
and therefore $ E_{WFE}(\psi) $ of order $ wN $, of order $ 1/N^2 $ close to \emph{pure states} (classical configurations with no superpositions) and therefore $ E_{WFE}(\psi) $ of order $ w $. We observe that for the macroscopic magnetization $ M(\psi)= N m(\psi) $ the situation will be opposite, that is it will be very small close to product and cat states while of order $ N $ close to  pure states.

In Math Appendix \ref{app:models no stat} we show that adding wavefunction energy to SQUIM 
(call the resulting model SQUIMWFE) does not change the conclusion 
about magnetization (namely, none at any finite temperature).

Namely, even under non-linear modifications, it is not possible to produce phase transitions.  Which we attribute to entropy swamping energy because of high dimensionality of the models:
\begin{quote}
	\centerline{\bf Proposition}\label{prop:1}	
	For distinguishable particles with Hamiltonian \eqref{eq:genwfe}, given a magnetic spin energy $ E(S) $ of order $ N $, e.g. Curie-Weiss or Ising,  in any dimension  for any finite temperature (except for $ T=0 $) there is no magnetization in the thermodynamic
	limit:
	\be\label{eq:mwithWFE}
	\lim_{N\to \infty}\,[\,\{\, |m(\psi)| > \epsilon\,\}\,]_{\beta} = 0,
	\ee
	\ni for any $\epsilon > 0$.
\end{quote}

\section{The SCWM with Wavefunction Energy}\label{sec:SCWwithWFE}

The classical Curie-Weiss (CW) model has energy:

\be
E_{CW} \= -\frac{1}{N}\, \left(\,\sum_{i=1}^N\,S_i\,\right)^2.
\ee

So the magnetic energy in a symmetric-\wf\, model associated to $\phi$ becomes:

\be
E_{CW}(\phi) \= - \frac{1}{N}\, \sum_{n=0}^N\,|\phi_n|^2\,\left(\,N-2n\,\right)^2.
\ee

A quick calculation gives:

\be
E_{CW}(\phi) \= -N\,\left\{\,m^2(\phi) + D(\phi)\,\right\},\label{qcal}
\ee

\ni where

\bar
\no m(\phi) &\=& \frac{1}{N}\,\sum\,|\phi_n|^2\,\left(\,N-2n\,\right);\\
\no D(\phi) &\=& \frac{1}{N^2}\,\left\{\,\sum\,|\phi|^2\,\left(\,N-2n\,\right)^2 \- 
\left[\,\sum\,|\phi_n|^2\,\left(\,N-2n\,\right)\,\right]^2 \,\right\}.\\
&&
\ear

In passing, we point out the curious fact, revealed by (\ref{qcal}), 
that $D(\phi)$ plays a role in a \wf\ model
with the standard energy.
We define:

\be
f \= N\,\beta\,\left\{\, 1 - m^2(\phi) - D(\phi) + N\,w\,D(\phi)\,\right\}.
\ee

\ni Here we have added a term ($N\beta$) to make $f\geq 0$ and incorporated \wf\ energy.
The intuition for the latter choice comes from the observation that, lacking that term,
$f$ can be small if {\em either} $m^2$ is large {\em or} $D$ is large; incorporating
the dispersion term with large enough $wN$, the last possibility should be surpressed.

As always, the model is defined by:

\bar
 [\,m^2\,]_{\beta} \= \int_{||\phi||^2 = 1}\,d\phi\,\exp\{ -f(\phi) \}\,m^2(\phi)/Z;\,\,\,\, Z \= \int_{||\phi||^2 = 1}\,d\phi\,\exp\{ -f(\phi) \}.
\ear

Intuition suggests we should investigate cases were $w$ is at least $1/N$; 
hence we define

\be
\omega \= N\,w;
\ee

\ni and we assume below that $\omega$ is a constant. This 
does not indicate a belief that $w$ actually scales with $N$; if wavefunction energy
exists, then $w$ is a constant of nature and does not scale. 
The role of the assumption is to avoid suppressing all superpositions, 
as would follow with fixed '$w$' 
in the mathematical limit of large $ N $ because of the factor $ N^2 $. 
But this limit does not exist in reality; 
so our assumption is just a mathematical stratagem to prove theorems.

In Math Appendix \ref{app:SCWwithWFE} we prove:

\begin{quote}
\centerline{\bf Theorem Three}
Let positive numbers $\omega$ and $\epsilon$ satisfy:

\bar
 && 1 < \omega < 4/3;\label{eq:rangeomega}\\
 && \epsilon < \frac{1}{4}\,\left(\, 1 + \sqrt{1 - 4\,r}\,\right)^2;\label{epineq}\\
&& r \= \frac{\omega - 1}{\omega}.
\ear

Then there is a positive number $\beta_c$ such that, for $\beta > \beta_c$,

\be
\lim_{N \to \infty}\,[m^2]_{\beta} > \epsilon.
\ee

\end{quote}
The bound $ 4/3 $ is  required for the application   of G\"artner-Ellis' theorem but  we expect that is has no meaning. 
 With the specified range for $\omega$, (\ref{epineq}) always holds if $\epsilon < 1/4$.

\section{Have We Chosen the Right Ensembles?\label{leonardo}} 

In previous sections, we constructed Gibbs ensembles where each state 
is weighted with a uniform measure on the sphere. 
{\em A priori}, we could consider other measures conserved by the Hamiltonian flows. 
E.g., we could consider modifications of the uniform one 
as $ d\phi F(\phi) $ where $ F(\phi) $ satisfies  $ \{F(\phi), H(\phi)\}=0 $, 
where $\{\cdot\}$ indicates Poisson bracket. 
In the quadratic case this will be equivalent to $[F,H]=0  $, where now $[\cdot]$ denotes
the  commutator. This reflects the fact that  there are many  conserved quantities 
other than the energy and the norm in linear QM (including the moduli of the wavefunction
component in the eigenstate-directions), 
showing that linear QM is far from being ergodic. 
 As discussed in  \cite{MP1}, Schr\"odinger's equation and 
the non-linear modifications proposed there are simply Hamilton's systems disguised; 
hence,  Liouville's theorem applies, implying that the base measure 
should be $ \prod_{S} d\psi(S) $.  The non-linear modification of \cite{MP1} 
of course will eliminate the  conserved quantities along eigenstates directions.
Proving the ergodicity for some such modification 
would qualify the uniform measure as the unique choice for a base measure. 
But  at the present time we can only demonstrate the 
existence of expanding and contracting directions in certain special cases, see \cite{MP2}.

An alternative ensemble, with the same base measure we used, could be constructed  
by adopting a Boltzmann factor that weights
 a state $ \psi $ taking into account the notion of possible microscopic states 
compatible with a macroscopic one. Concerning the quantum theory, 
this notion can be tracked back to Einstein, \cite{einstein}, 
in a paper about how to define quantum entropy.  We think
that if the introduction of this modification has a meaning, it should be related to having
considered indistinguishable particles jointly with a lack of spatial coordinates, which would
add a degeneration on energy levels. For example confining each spin in a spatial potential, we
would have many levels for each spin and correspondingly many ways to arrange
n spins down.
In the SCWM case, the proposal would be to replace in $ \prod_{n} d\phi(n) $  
with  $ \prod_n  [C(N,n)]^{|\phi_n|^2}d\phi_n $, where $ C(N,n) $ 
is the number of states with $ n $ spins down.  
In this way, a wavefunction acquires a large combinatorical weight 
if amplitudes are concentrated on highly-degenerated components.

Calling  $ \mathcal{H} $ the Hilbert space of wavefunctions  
of norm one decomposed as $ \mathcal{H}=\underset{n}{\oplus}\mathcal{H}_n$, 
where $ \mathcal{H}_n $ is the subspace of the wavefunctions with 
quantum number $ n $ and dimension $ \text{dim}\mathcal{H}_n:=C(N,n) $, 
the proposed measure can be  written also as 

\begin{equation}\label{eq:Boltz measure}
 \prod_n  [\dim \mathcal{H}_n]^{|\phi_n|^2}d\phi_n= e^{ [\log\dim \mathcal{H}](\phi)}\prod_n d\phi_n,
\end{equation}

\ni with $[\log\dim \mathcal{H}](\phi)= \underset{n}{\sum} |\phi_n|^2  
\log (\text{dim}\mathcal{H}_n )$ and it is verified
$ \{[\log\text{dim} \mathcal{H}] (\phi), H_{SCWM}(\phi)\}=0 $.

Whether such a modified
measure is still conserved by the Hamiltonian flow, with or without WFE,
will depend on the precise details of the dynamics. (One author did introduce
a model of dynamics for a small number of qu-bits, \cite{MP2}, adopting an {\em ad hoc}
form of ``kinetic energy"; but this construction cannot be regarded as modeling a magnet.)

An {\em a priori} weight associated to a single wavefunction is questionable. 
But at the moment we can not exclude  a different measure than the uniform one on the state space. 
Let's reflect on how this ensemble would introduce an analogous notion 
to the classical  microcanonical  entropy of the usual spin models.
Introducing the energy and entropy densities respectively
\begin{equation*}\label{eq:e}
e(\phi)=  \langle\psi|e|\psi\rangle, \text{ where }  e(n)=-m_N^2(n)=- \left(1-\frac{2n}{N}\right)^2,
\end{equation*} 
\begin{equation*}\label{eq:s}
s(\phi)=\frac{1}{\beta}\langle \psi | s|\psi\rangle, \text{ where }s(n)= \frac{1}{N }\log C(N,n).
\end{equation*}

\ni By Sterling, for  large $ N $,  $ s(n) $   is approximated by
\begin{equation*}\label{eq:entropym}
s(m_N(n))=-\frac{1-m_N(n)}{2}\log\frac{1-m_N(n)}{2}-\frac{1+m_N(n)}{2}\log\frac{1+m_N(n)}{2}.
\end{equation*}

\ni We also introduce
\begin{equation*}\label{eq:frendens}
f^\beta(\phi):= e(\phi)-s(\phi).
\end{equation*}
The partition function becomes
\begin{align}\label{eq:symCW2}
 \frac{1}{S_N}\int_{\|\phi\|=1} e^{ \log\dim \mathcal{H}_n(\phi)} d\phi \,\,\exp(-\beta E(\phi))\approx \frac{1}{S_N}\int_{\|\phi\|=1}d\phi \,\, e^{ -\beta N f^\beta(\phi)}.
\end{align}
Since we are interested in making $ N $ large,
 we used the approximation $ \ \frac{1}{\beta N}\log\dim \mathcal{H}_n(\phi)\approx  s(\phi) $ 
and, neglecting the errors, we arrive at  
\begin{align}\label{eq:symCW3}
Z_N = \frac{1}{S_N}\int_{\|\phi\|=1} d\phi \,\, e^{ -N\beta[ -(m(\phi))^2- s(\phi) -   D(\phi)]}.
\end{align}

\ni The extra factor $ e^{\beta Ns(\phi)} $  can be thought of as trying to restore a classical picture of a competition between internal energy and entropy due to high degeneracy of disordered macrostates.
Anyway, since the role of this factor is giving large weights to some states of zero magnetization the WFE will be still necessary to define magnetic models.

The way to recover the usual ensembles 
may be to introduce $ wN^2\,D $ with $ w $ constant 
(i.e. without keeping $ \omega=wN $ constant) in the thermodynamic limit.
In this case, one expects that the measure will concentrate on the set $ \{\phi: D(\phi)=0\} $. 
Conceivably, we might arrive at the classical ensembles 
even in models without exchange symmetry, as considered in section \ref{sec:SQUIM}.
But, apart from conflicting with 
a fundamental fact of Quantum Mechanics, i.e. 
that particles are indistinguishable and so wavefunctions must have symmetries, 
the ensemble will not magnetize in the thermodynamic limit even with suppression of superpositions, 
as already discussed at the end of section  \ref{sec:ModwithWFE}. 
We mention  this is the route so far followed by various authors,
\cite{Bloch,Jona-Presilla,lebowitz,ES1,ES2},  
where they reduced the ensembles on the sphere to the usual ones by
 introducing delta measures on eigenstates by different reasoning. 
 The concentration we expect to be true for indistinguishable particles.

We guess that the uniform measure is still the right choice. The introduction of the
Boltzmann factor appears a bit artificial, but it seems relevant when including $ wN^2 D  $ with
$ w $ constant, as otherwise the internal energy would have no competition, assuming to the
measure concentrates on pure states. But we suggested there could be other meaningful pictures.

\section{Discussion}\label{sec:discussion}
 
A key element in the proof of Theorem One is that the first term in equation (\ref{yyeq})
vanishes, because of the overall spin-flip invariance of the model. 
This assumes ``free" boundary conditions. If we had enforced positive boundary conditions, 
this term would contribute to the magnetization. 
Does it nevertheless tend to zero as $N \to \infty$? Or might these
\Schist\ quantum models differ in behavior from classical models, in the sense that the system
retains a ``memory" of boundary conditions above some critical temperature, even though 
with free boundary conditions no magnetized states would appear in that limit? 
(We doubt it is so since the entropy still comes with that high dimensionality.)

On the other hand, in section \ref{leonardo}, we questioned the use of the uniform ensemble.
Might our negative results about SQUIM reflect a failure to adjust for the high-dimensionality,
and hence high entropy, incurred by replacing classical configurations by wavefunctions?
In which case phase transitions for combininations of distinguisable systems 
(e.g., ``qubits" located in separate devices) might still be on the table. 
We leave this question for further investigation.

We observe that the proof of Theorem One would proceed unchanged 
if we considered the mean field instead of nearest neighbors;
therefore,  we might interpret the necessity to consider  
exchange symmetry if we hope to have a thermodynamics that includes phase  transitions 
as another justification of the usual quantum rules for combining indistinguishable systems. 

Our models are oversimplifed. Would similar results arise 
for a SQUIM model with three-dimensional rotational symmetry, 
e.g., in a \Schist\ version of the Heisenberg model?

Concerning the proofs of the theorems: we cannot, unfortunately, amalgamate three into one,
as they used distinct techniques.
For Theorem One we compared the SQUIM to an exactly solvable model which lacked a phase transition.
Theorems Two and Three relied on the \Gartner-Ellis theorem of Large Deviations theory, 
but used it for different purposes.
This stemmed from the fact that, with the usual energy expression,
the exponent in the probability measure is quadratic so we have a Gaussian
integral we can compute explicitly. But adding wavefunction energy spoils this 
convenient scenario, as the exponent is quartic; so in the proof of Theorem Three we utilized
a result about support of a measure in a limit (the ``Concentration Lemma", see Math Appendix \ref{app:lemmas}) 
and inequalities derived from factoring the
energy expression.

Assuming the validity of Theorem Two (which claims that SCW with the usual energy does not
spontaneously magnetize), our results suggest that blocking magnetic cats is related
to phase transitions, presumably because states which are broad superpositions 
tend to have low magnetizations. It might be objected that 
the non-linear modification of \Sch's\ equation by addition of WFE
 introduced in \cite{MP1} was proposed 
to eliminate spatial dispersion and not for sums of discrete ``spins" as we examined here.  The straightforward
generalization of \eqref{eq:WFEspin} to models with spatial coordinates is $ O_i = L_i + S_i $ in \eqref{eq:genwfe}. 

Perhaps the two main tasks for  future investigations (which appear already very hard) are:  
improving the ability to compute spherical integrals, in 
the hope to derive the functional dependence of magnetization on temperature;
and generalizing the model to the case $ \{-1,0,1\} $ for symmetric wavefunctions and to the case $ \{-1,1\} $  for antisymmetric wavefunctions.  We observe that these cases will automatically add  degeneration to energy levels.

Also it might worth to  consider  $ N$-levels for each spin, namely a pictorial way for  replacing the discrete lattice with a confining potential for each spin  and studying the limit for $ wN^2 $, this might help an understanding of Boltzmann entropy from a quantum perspective.

\section*{APPENDICIES}

Here we collect in several appendicies the various computational lemmas
 we used in the proofs that follow. 

\appendix

\section{Math Appendix: Some Useful Lemmas}\label{app:lemmas}

Let $S$ be a compact manifold without boundary, f a real-valued function on $S$, $dx$ a finite
measure on $S$. Without loss of generality, we can take 

\be
\int_S\,dx\, \= |S| \= 1,
\ee

\ni and assume $f \geq 0$.

Let, for any bounded $g$ on $S$:

\bar
 [g] \= \int_S\,dx\,\exp\{ - f(x)\,\}\,g(x)/Z;\,\,\,\, Z \= \int_S\,dx\,\exp\{ - f(x)\,\}.
\ear

\begin{quote}
\centerline{\bf Concentration Lemma} 

Let there be two open subsets of $S$, 
called $U$ and $V$, and three positive numbers
$\alpha$, $\eta$, and $\mu$ such that 
\bar
\no &&\hbox{(A)}\ph\ph V \subset U;\\
\no &&\\
\no &&\hbox{(B)}\ph\ph f(x) \leq \eta, \ph\hbox{for}\ph x \in V; \\
\no &&\\
\no &&\hbox{(C)}\ph\ph f(x) \geq  \alpha, \ph\hbox{for}\ph x \notin U; \\
\no &&\\
\no && \hbox{(D)}\ph\ph |V| \geq \,\mu.\\
&&
\ear

Then: 

\be
[g] \= \frac{R+\xi}{1 + \zeta},
\ee

\ni where

\bar
\no R &=& \int_U\,dx\,\exp\{ - f(x)\,\}\,g(x)/Z_U ;\\
\no Z_U &=& \int_U\,dx\,\exp\{ - f(x)\,\};\\
\no|\xi| &\leq& e^{- \alpha}\,e^{\eta}\,\mu^{-1}\,||g||;\\
\no|\zeta| &\leq& e^{- \alpha}\,e^{\eta}\,\mu^{-1}.\\
&&
\ear
\end{quote}

Here $||g||$ denotes the supremum norm of $g$ on $S$. 
Note that $R \in \hbox{span}\{\,g(x):\,x \in U\,\}$.

The intuition behind this lemma is the following. Suppose we are expecting the measure with
density $\rho = e^{-f}/Z$ to be close to a delta function. Then this density must have a spike at
the minimum of $f$, call it $\rho_*$, and the ratio of $\rho_*$ to values of $\rho$ far 
away from the spike should
tend to infinity. But we still 
won't get a delta function if the spike occupies a very tiny part of the
manifold (imagine it as an infinitely-thin needle) 
because it will contribute little to the integral. 
Hence the roles of the sets $U$ and $V$ and the 
bounds on $f$ which prevent this ``needle-like" behavior. 
The idea is that $V$ is a small neighborhood of the global
minimum of $f$. The minimum may not occur at a single point, but on a subset. 
The volume $|V|$ should not be 
so small as to put a large factor in $\xi$ and $\zeta$; while $\alpha > \eta$. Then $\xi$ and
$\zeta$ should tend to zero and the measure concentrates on the set $U$.

We note that the ``balance of energy and entropy" business is contained in the difference
$\alpha - \eta$ and in $\mu$, measuring how $f$ increases 
compared with the volume of the manifold that requires.

Next consider an integral of form:

\be
\int_B\,\exp\{\,-f(\phi)\,\}\,d\phi,
\ee

\ni where $d\phi$ stands for a probability distribution on
 a compact manifold with continuous density, 
$B$ is a regular subset (non-empty interior and boundary of $d\phi$-measure zero)
 and $f$ is a continuous 
function on $B$ with values in $[0,c]$ and level sets of zero $d\phi$-measure. 
Then we have the representation:

\begin{quote} 
\centerline{\bf Lemma (Useful Integral Formula, UIF)} 
\be
\int_B\,\exp\{\,-f(\phi)\,\}\,d\phi = e^{-c}|B| \+ c\,\int_0^1\,dx\, e^{-c\,x}\,F(x),
\ee
\ni where
\be
F(x) \= |\,\left\{\,f(\phi) \leq c\,x;\ph \phi \in B\,\right\}\,|.
\ee
\end{quote}
\ni Here, $|\cdot|$ denotes the volume (``area") of the subset $B$.

{\bf Proof}:

\bar
\no && \int_B\,\exp\{\,-f(\phi)\,\}\,d\phi \geq \sum_{k=0}^K\,e^{-kc/K}\,\left| \,\left\{\,\frac{(k-1)c}{K}\,< f(\phi) \leq \frac{kc}{K};
\,\,\phi \in B\,\right\}\right|\,\\
\no && \sum_{k=0}^K\,e^{-kc/K}\,\left\{\,F\left(\frac{k}{K}\right) - F\left(\frac{k-1}{K}\right)\,\right\} \approx \frac{1}{K}\,\sum_{k=0}^K\,e^{-kc/K}\,F'\left(\,\frac{k}{K}\,\right).\\
&&\label{calc}
\ear

The last line tends, as $K\to\infty$, to:

\be
\int_0^1\,dx e^{-cx}\,F'(x).
\ee

We can get an identical upper bound in the limit $K\to\infty$ by substituting 

\be
e^{-(k-1)c/K} \= e^{-kc/K}\,e^{c/K}
\ee

\ni in the calculation leading to (\ref{calc}).

Now the Lemma follows from integration-by-parts, using that $F(0) = 0$ and $F(1) = |B|$. QED 

\section{Math Appendix: A Curious Computation: the Average Magnetization at T $= \infty$}\label{app: computation}

\def\cE{{\cal{E}}}
\def\dphi{{\int\,d\,\phi}}
\def\sumnN{{\sum_{n=0}^N}}
\def\summN{{\sum_{m=0}^N}}
\def\phinsq{{|\phi_n|^2}}
\def\phimsq{{|\phi_m|^2}}
\def\oA{{\frac{1}{A_N}}}

In this section we ask: what is the magnetization per spin at T = $\infty$, meaning $\beta = 0$?
It should certainly be zero, which we prove here; the calculation will also provide
some tools useful further on.

We adopt symmetric \wfs, which we denote by $\phi_n$, for $n=0,...,N$. 
Here $n$ denotes the number of ``up" spins. To avoid factors of 1/2 below 
we take our spins to have values $\pm1$. 

In the following the expression:

\be
\dphi
\ee

\ni will be understood to be an integral over the normalized measure on the sphere,
i.e.,

\be
\dphi = \oA\,\int_{||\phi|| = 1}\,\prod\,d\phi_n.
\ee

\ni where $A_N$ is the 'area' of the 2N-sphere (the hypersphere in $R^{2N}$).

We can write:
\bar
\no \left[\,m^2(\phi)\,\right]_{\infty} &\=& \left[\,\left(\,\frac{M(\phi)}{N}\,
\right)^2\,\right]_{\infty}= \dphi\,\left\{\,\sumnN\,|\phi_n|^2\,g_n\,\right\}^2;\\
&&
\ear
\ni where
\be
\no g_n \= 1 - 2n/N.
\ee
Developing:
\bar
\no \left[\,m^2(\phi)\,\right]_{\infty} &\=& \sumnN\,\summN \dphi\,\phinsq\,\phimsq\,g_n\,g_m= c_1\,\sumnN\,g_n^2 + c_2\,\sum_{n \neq m}^N g_n\,g_m,\\
&&\label{msqeq}
\ear
\ni where
\bar
 c_1 \= \oA\,\dphi\,|\phi_1|^4;\,\,c_2 \= \oA\,\dphi\,|\phi_1|^2\,|\phi_2|^2.
\ear
By adding and subtracting a term in (\ref{msqeq}) and using that
\be
\sumnN\,g_n = 0,
\ee
\ni we can write:
\be
\left[\,m^2(\phi)\,\right]_{\infty} \= (c_1 - c_2)\,\sumnN\,g_n^2.\label{msqform}
\ee
The sum can be evaluated from formulas:
\bar
\no \sumnN\,g_n^2 &\=& N + 1 - \frac{4}{N}\,\sumnN\,n + \frac{4}{N^2}\,\sumnN\,n^2\\
\no &\=& N + 1 - \frac{4}{N}\,N(N+1)/2 + \frac{4}{N^2}\,N(N+1)(2N+1)/6\\
\no &\=& \frac{1}{3}\,N \+ \hbox{smaller order}.\\
&&
\ear
We conclude that
\be
\lim_{N \to \infty} \frac{1}{N}\,\sumnN\,g_n^2 \= 1/3.\label{gnlimit}
\ee
Putting in factors of $N$ and $1/N$ in (\ref{msqform}) we deduce that we must consider
the limit:
\be
\lim_{N \to \infty} N\,(c_1 - c_2).
\ee
By the same tricks but replacing $g_n$ by one, we find:
\be
1 = (N+1)\,c_1 \+ c_2\,N\,(N-1),
\ee
\ni from which we conclude that $c_1$ = O($1/N$) and $c_2$ = O($1/N^2$), so the issue becomes
evaluating:
\be
\lim_{N \to \infty} N\,c_1.
\ee

Informally, this limit should be zero, since $|\phi_n|^2 \approx 1/(N+1)$ on average,
so $c_1$ should also be O($1/N^2$).
To prove this rigorously, we resort to the polyspherical coordinate system: 
let $\xi$ be a point on the (n-1)-sphere (meaning
the sphere in $R^n$).
Write\footnote{V\&K, 9.19}
\be
\xi \= \eta\,\sin(\theta) \+ \zeta\,\cos(\theta),
\ee
\ni where $\eta$ lies on the (s-1)-sphere and $\zeta$ lies on the (n - s - 1)-sphere. Then:
\be
d\xi \= b_n\,\sin^{s-1}(\theta)\,\cos^{n - s - 1}(\theta)\,d\eta\,d\zeta\,d\theta,
\ee
\ni where 
\be
b_n \= \frac{2\Gamma(n/2)}{\Gamma(s/2)\,\Gamma([n-s]/2)}.
\ee
We take $s = 2$ and $\eta$ as the first component of $\xi$ and can therefore write:
\bar
\no \int\,d\xi\,|\xi_1|^4 &\=& b_n\,\int_0^{\pi/2}\,d\theta\,\sin(\theta)\,\cos^{n-3}(\theta)
\,\int\,d\eta\,\sin^4(\theta)\,|\eta|^4\,\int\,d\zeta\\
\no &\=& b_n\, \int\,d\eta\,|\eta_1|^4\,
\int_0^{\pi/2}\,d\theta\,\sin^5(\theta)\,\cos^{n-3}(\theta).\\
&&
\ear
\ni Making the substitution $u = \cos(\theta)$ the above integrand is a polynomial
and the integral works out to be,
substituting $n = 2N$:
\be
\frac{8}{n(n^2 - 4)} \= \frac{1}{N(N^2 - 1)};
\ee
\ni also, the prefactor comes out:
\be
b_{2N} \= 2\,(N-1),
\ee
\ni which proves the assertion about the order of $c_1$ and that the average magnetization
at infinite temperature is zero in the thermodynamic limit.
 
This theorem can also be proved using LD theory but we omit it as covered in other Appendicies.

\section{Math Appendix: models without exchange symmetries} \label{app:models no stat}

For the computations of this section we add to the energy an external field with a constant 
$\lambda$, incorporate the factor of $\beta$, and take the positive version of the interaction
energy (which merely multiplies $Z$ by a factor): 

\bar
 E(\psi;\lambda) \= \beta\,
\sumS\,\sum_{i,j;\, n.n.}^N\,\left(\,S_i - S_j\,\right)^2 + \lambda\,M(\psi);\,\,\,\,M \= \sumS\,\sum_i\,S_i.
\ear
\ni Here and below ``$\sum_S$", resp. ``$\prod_S$", is shorthand for the sum,
resp. product, over all configurations $\{\,S_1= \pm 1/2,...,S_N = \pm1/2\,\}$.
we will also write
\be
E(\psi;\lambda) \= \sumS\,\ESl.
\ee

The starting point is to replace the spherical integral by a Gaussian integral,
preserving the definition of the model by projecting $\psi$ onto the sphere.
Hence we can write:

\bar
\no Z_N \= c_N\,\int\,\prod_S\,d\psi_S\,\exp\left\{\,- ||\psi||^2/2 - E(\psi/||\psi||;\lambda)/2\,
\right\};\,\,\,\
||\psi||^2 \= \sumS.
\ear

\ni Here we have introduced a factor of 1/2 before the energy, 
which simplifies some calculations below.

The basic idea is to note that, although the integrand is singular at $\psi = 0$, 
because the components of $\psi$ are i.i.d. $N(0,1)$, that value is improbable; indeed,  
\be
\sumS \approx a_N = 2\,2^N.
\ee

Therefore we define a simpler model by replacing the worrisome 
sum in the integrand by a constant:
\be
\hatZ_N \= c_N\,\int\,\prod_S\,d\psi_S\,\exp\left\{\,- ||\psi||^2/2 - E(\psi;\lambda)/2a_N\,
\right\};
\ee

\ni which we will dub the ``Exactly Solvable Model'' (ESM), as we are reduced to computing
a Gaussian integral.
We have:
\be
\hatZ_N \= c_N\,\prod_S\,\int\,d\psi_S\,\exp\left\{\,- ||\psi||^2/2\sigma_S^2\,
\right\};
\ee
\ni where 
\be
\sigma_S^2 \= \left\{1 + \,E_S/a_N\,\right\}^{-1},
\ee
\ni so from the basic Gaussian integral in two-dimensions we find:
\be
\hatZ_N \= \prod_{S} \, 
\frac{1}{1 + E_S/a_N}.\label{formula}
\ee
Next we take two derivatives with respect to $\lambda$, denoted with primes, divide
by $\hatZ_N$ and evaluate at $\lambda = 0$:

\bar
\ \frac{\hatZ_N^{''}}{\hatZ_N}|_{\lambda = 0} \=
\left\{\,\frac{1}{a_N}\,\sum_S\, \frac{M_S}{(1+E_S/a_N)}\,\right\}^2 \+   \frac{1}{a_N^2}\,\sum_S\,\frac{M_S^2}{(1+E_S/a_N)^2}.\label{yyeq}
\ear
\ni The first term is zero by the spin-flip symmetry ($M_S$ changes sign but since $E_S$
is quadratic in $S$ it doesn't). Then, observing the prefactors of $1/a_N^2$, the whole
thing tends to zero as $N\to\infty$ even without a factor of $1/N^2$, 
so there is no phase transition in the ESM.

Finally we must estimate the difference in magnetizations in the two models.
We can rewrite $Z_N$ as:
\be
Z_N \= c_N\,\int\,\prod_S\,d\psi_S\,\exp\left\{\,- ||\psi||^2/2 - E(\psi)/2a_N + \chi_N\,\right\},
\ee
\ni where
\be
\chi_N \= \half\,\frac{E(\psi)}{||\psi||^2}\,
\left\{\,\frac{||\psi||^2}{a_N} - 1\,\right\}.
\ee
Also, define:
\be
\xi_N \= \frac{1}{4}\,\left\{\,\frac{M(\psi)}{||\psi||^2}\,\right\}^2\,
\left\{ 1 - \,\Big(\,\frac{||\psi||^2}{a_N}\,\Big)^2 \,\right\}.
\ee
By expanding the exponential we can write
\bar
\no Z_N &\approx& \hatZ_N \+ 
 c_N\,\int\,\prod_S\,d\psi_S\,\exp\left\{\,- ||\psi||^2/2 - E(\psi)/2a_N\,\right\}\,\chi_N;\\
\no Z_N^{''} &\approx& \hatZ_N^{''} \+ 
 c_N\,\int\,\prod_S\,d\psi_S\,\exp\left\{\,- ||\psi||^2/2 - E(\psi)/2a_N\,\right\}\,
\xi_N \+\\
\no && c_N\,\int\,\prod_S\,d\psi_S\,\exp\left\{\,- ||\psi||^2/2 - E(\psi)/2a_N\,\right\}\,
\Big(\,\frac{M(\psi)}{||\psi||^2}\,\Big)^2\,\chi_N;\\
&&
\ear
Hence, expanding to first order in the small quantities $\chi_N$ and $\xi_N$,
\bar
\no \frac{Z_N^{''}}{Z_N} &\approx&
 \frac{\hatZ_N^{''}}{\hatZ_N} \+  c_N\,\int\,\prod_S\,d\psi_S\,\exp\left\{\,- ||\psi||^2/2 - E(\psi)/2a_N\,\right\}\,
\xi_N/\hatZ_N \-\\
 \no && c_N\,\int\,\prod_S\,d\psi_S\,\exp\left\{\,- ||\psi||^2/2 - E(\psi)/2a_N\,\right\}\,
\Big(\,\frac{M(\psi)}{||\psi||^2}\,\Big)^2\,\chi_N/\hatZ_N \-\\
\no && \frac{\hatZ_N^{''}}{\hatZ_N^2}\,c_N\,\int\,\prod_S\,d\psi_S\,
\exp\left\{\,- ||\psi||^2/2 - E(\psi)/2a_N\,\right\}\,
\chi_N. \\ 
&& \label{expansion}
\ear

Let's treat the third term first. Noting that

\bar
 \Big(\,\frac{M(\psi)}{||\psi||^2}\,\Big)^2 \leq (1/4)\,N^2;\,\,\,\,
\frac{E(\psi)}{||\psi||^2}\leq (1/4)\,N.
\ear
Applying Cauchy-Schwarz we can bound the term by:
\bar
\no && N^3/16\,\left\{\,c_N\,\int\,\prod_S\,d\psi_S\,
\exp\left\{\,- ||\psi||^2/2 \right\}\,
\left\{\,\frac{||\psi||^2}{a_N} - 1\,\right\}^2
\right\}^{1/2}\, \times\\
\no && \left\{\,c_N\,\int\,\prod_S\,d\psi_S\,
\exp\left\{\,- ||\psi||^2/2 - \,E(\psi)/a_N\,\right\}\,
\right\}^{1/2}/\hatZ_N.\\
&&
\ear 
By standard CLT calculations, the first factor in curly brackets tends to zero, 
at rate $1/\sqrt{a_N}$. 
Using (\ref{formula}) the second factor can be rewritten as:
\be
\left\{\,\prod_S\,\frac{1 + 2E_S/a_N + E_S^2/a_N^2}{1 + 2E_S/a_N}\,\right\}^{1/2},
\ee

\ni which is easily seen to be bounded (and in fact tends to one).
In treating the second term we encounter instead the quantity:

\bar
\no && \left\{\,c_N\,\int\,\prod_S\,d\psi_S\,
\exp\left\{\,- ||\psi||^2/2 \right\}\,
\left\{\,\Big(\,\frac{||\psi||^2}{a_N}\,\Big)^2 - 1\,\right\}^2
\right\}^{1/2}\\
&&
\ear

\def\cE{{\cal{E}}}
This term can also be shown to tend to zero. E.g., writing ``$\cE$" for the Gaussian integral,
factoring the difference of squares and making another Cauchy-Schwarz, we have to estimate:

\bar
\no \cE\,\Big(\,\frac{\sum\,|\psi_S|^2}{a_N} - 1\,\Big)^4 &\=& 
 \cE\,\Big(\,\frac{1}{a_N}\,\sum\,\left\{\,|\psi_S|^2 - 1\,\right\}\,\Big)^4\\
\no &\=& \frac{1}{a_N^4}\,\sum_{S,S',S'',S'''}\,\cE\,\left\{\,|\psi_S|^2 - 1\,\right\}
\cdot\cdot\cdot
\left\{\,|\psi_{S'''}|^2 - 1\,\right\}\\
\no &\=& \frac{6}{a_N^4}\,\left\{\,\sum_S\cE\,\Big(\,|\psi_S|^2 -1\,\Big)^2\,\right\}^2 \+
 \frac{4}{a_N^4}\,\sum_S\cE\,\Big(\,|\psi_S|^2 -1\,\Big)^4\\
&&
\ear

\ni which is O$(a_N^{-2})$.

We also have to estimate 

\be
\cE\,\left\{\,\frac{||\psi||^2}{a_N} + 1\,\right\}^4,
\ee

\ni which by the same tricks is seen to be bounded.
Recalling that
\be 
\frac{\hatZ_N^{''}}{\hatZ_N}\,\to\,0,
\ee
\ni the other term in (\ref{expansion})
 is even easier to treat. 

Since moments of Gaussians increase more slowly than a factorial, 
the sum of the other terms is dominated by a convergent series and of smaller order. QED

Now consider adding wavefunction energy:
 
\bar
\label{eq:EWFE} E_{WFE}(\psi) &\=& w N^2\,D(\psi);\\
\no D(\psi) &\=& \frac{1}{N^2}<\psi|\,\big(\,\sum\,S_i - <\psi|\,\sum\,S_i\,|\psi>\,\big)^2\,|\psi>\\
\no &\=& \frac{1}{N^2}\left[ \sum_{i,j} <\psi|\,S_i\,S_j\,|\psi> - \left\{\,\sum_i\, <\psi|\,S_i\,|\psi>\,\right\}^2\right].\\
&&\label{Deq}
\ear
Let's switch to the Gaussian version, divide $\psi$ by $||\psi||$,
 and replace $||\psi||^2$ by $a_N$. 
Treating the second term in the last line above, define
\be
X_{\pm,i} \= \frac{1}{2a_N}\,\sum_{\hat{S}}\,|\psi|^2(S_1,...,\pm\half,...,S_N),
\ee
\ni where ``$\sum_{\hat{S}}$" means summation over the spin configurations 
with $S_i$ held at the fixed value indicated. Then we can write:
\be
<\psi|\,S_i\,|\psi> = \frac{1}{a_N}\,\sum_S\,|\psi(S)|^2\,S_i \= X_{+,i} - X_{-,i}.
\ee

Note that, using $\cE$ for expectation over
the Gaussian distribution:

\bar
\no \cE\,\big(\,X_{+,i} - X_{-,i}\,\big) &\=& 0;\\
\no \cE\,\big(\,X_{+,i} - X_{-,i}\,\big)^2 &\=& 
\cE\,\big(\,X_{+,i} - \cE\,X_{+,i}\,\big)^2 +
\cE\,\big(\,X_{-,i} - \cE\,X_{-,i}\,\big)^2;\\
\no &&
\ear 

\ni and

\be
 \cE\,\big(\,X_{+,i} - \cE\,X_{+,i}\,\big)^2 \= \cE\,\left\{\,
\frac{1}{2a_N}\,\sum_{\hat{S}}\,\left[\,|\psi|^2(S_1,...,\half,...,S_N) - 2\,\right]\,\right\}^2.
\ee

Noting the average of i.i.d. mean-zero \rvs\, this last is O$(1/a_N)$.

The last term in (\ref{eq:EWFE}) is a sum of $N$ terms which are not independent but all of
the order just computed. By a simple Jensen inequality, the sum is at most $N$ times this order.
We conclude that the whole term is negligible. 

The first term is of form of a mean-field model, still quadratic in $\psi$. 
Hence it can be added to the usual magnetic energy. The bound on $E_S$ increases to $N^2$,
but doesn't affect the argument. 
For the comparison with the actual SQUIMWFE, we encounter the expression

\be
 \frac{1}{4}\,\left\{\,\frac{M(\psi)}{||\psi||^2}\,\right\}\,
\left\{\,\Big(\,\frac{||\psi||^2}{a_N}\,\Big) - 1\,\right\},
\ee

\ni which can be treated as before.
Thus the argument of the quadratic case goes through and yields 
that the SQUIMWFE has no magnetization
at finite temperature.

\section{Math Appendix: SCW without wavefunction energy}\label{app:SCWnoWFE}

The idea is to determine whether 

\be
\lim_{N\to \infty}\,\int_{m(\phi) \geq \epsilon}\,d\phi\,\exp\{\, -\,\beta\,E_{\hbox{CW}}\,\}/Z =  0,
\ee

\ni or not, where 

\be
Z \= \int\,d\phi\,\exp\{\, -\,\beta\,E_{\hbox{CW}}\,\}.
\ee

\ni and similarily with $\{m(\phi) \geq \epsilon\}$ replaced by $\{m(\phi) \leq - \epsilon\}$.
We can rewrite the integrals as:
\bar
 \int_{m(\phi) \geq \epsilon}\,d\phi\,\exp\{\, -f\,\}/Z, \,\,\, Z = \int\,d\phi\,\exp\{\, -f \,\},
\ear
\ni where
\bar
f = \beta\,N\,\left(\,1 - \sum\,|\phi_n|^2\,g_n^2\,\right);\,\,\,\, g_n = 1 - 2\,n/N.
\ear

\ni where as usual we have added a term so that $0\leq \,f\,\leq \beta\,N$. 

To compare these integrals in numerator and denominator, we turn to the UIF, setting
$B = \{m(\phi) \geq \epsilon\}$ in the numerator, and $B$ equal to the whole sphere in
the denominator. Thus for the numerator we will have to estimate:

\be
|\,\{ \beta\,N\,\left(\,1 - \sum\,|\phi_n|^2\,g_n^2\,\right) \leq \beta\,N\,x;\, 
\sum |\phi_n|^2\,g_n \geq \,\epsilon\,\}\,|,
\ee

\ni which, replacing wavefunctions on the sphere by i.i.d. Gaussians, equivalently: 

\be
\Plb (1-x)\,\sum \chi_n^2 \leq \sum\,\chi_n^2\,g_n^2; \,\sum\,\chi_n^2\,g_n \geq 
\epsilon\, \sum \chi_n^2 \Prb,
\ee

\ni (Since $\phi_n$ has two components, the sums are now over 2N+1 rather than N+1
indices, with the $g_n$ repeated; but as we are taking the limit as $N\to \infty$ we don't bother 
with factors of two everywhere.) 
\ni which is the same as writing (introducing factors of $1/N$ for later purposes):

\be
\Plb 1/N\, \sum \chi_n^2 \,(1 - x - g_n^2) \leq 0; 1/N \,\sum\,\chi_n^2\,(\,g_n - \epsilon\,) 
\geq 0 \Prb.\label{setB}
\ee

To motivate the appeal to the \Gartner-Ellis theorem, let
$P_{2;N}(x)$ stand for the probability of the set appearing above, 
and $P_{1;N}(x)$ for the probability with the second restriction dropped. 
We are interested in the ratio:

\be
p_N(x) = \frac{P_{2;N}(x)}{P_{1;N}(x)},
\ee

\ni which has the interpretation of the conditional probability that the magnetization
is greater than $\epsilon$, given a bound `$x$' on the energy. We expect that

\be
\lim_{N \to \infty} p_N(x) = 0.
\ee

Then, integrating over
$x$ (as in the UIF) we shall arrive at the result. Noting that the events considered
have the form of averages of random variables whose means lie outside the
indicated bounds, we expect large-deviations asymptotics for both, and so it is natural
to consider

\bar
 \lim_{N\to \infty} - \frac{1}{N} \log\,p_N(x)  \= \lim_{N\to \infty} - \frac{1}{N} \log\,P_{2,N}(x)  \+ \lim_{N\to \infty}  \frac{1}{N} \log\,P_{1;N}(x) 
\ear

\ni and to treat the two terms separately and then compare.

To implement the \Gartner-Ellis procedure, we introduce the random vector with two components:

\be
Y_N \= \left(\, 
1/N\, \sum \chi_n^2 \,(1 - x - g_n^2); \,1/N\,\sum\,\chi_n^2\,(\,g_n - \epsilon\,)\,\right),
\ee

\ni so we are interested in 

\be
\Plb Y_N \in \left[\,\hbox{Upper left quadrant of $R^2$}\,\right]\,\Prb.
\ee

To apply G-E we must compute:

\bar
\ && c(\theta_1,\theta_2) \= \lim_{N\to \infty}\, \frac{1}{N}\,\log\,{\cal E}\,\exp\{\,N\left(\,Y_{N;1}\,\theta_1 +
Y_{N;2}\,\theta_2\,\right)\,\} \=\\ 
\no && \lim_{N\to \infty}\,\frac{-1}{2N}\,
\sum\,\log\{\,1 - 2\,\left[\,\theta_1\,(1-x-g_n^2) + \theta_2\,
(g_n - \epsilon)\,\right]\,\} \=\\
\no && - \half\,\int_{-1}^1\,dy\, 
\log\{\,1 - 2\,\left[\,\theta_1\,(1-x-y^2) + \theta_2\,
(y - \epsilon)\,\right]\,\} \= - \half\,\int_{-1}^1\,dy\, 
\log\,q(y);\\
&&
\ear
\ni where 
\bar
\no q(y) &\=& 1 - 2\,\left[\,\theta_1\,(1-x-y^2) + \theta_2\,
(y - \epsilon)\,\right]\= 2\,\theta_1\,y^2 - 2\,\theta_2\,y + b,\\
&&
\ear
\ni where $b$ is a constant depending on the theta's. 
\def\twoth{{\theta_1,\theta_2}}

We must first determine the region in the $(\twoth)$ plane in which $c(\twoth) < +\infty$.
Defining
\be
 h(y) \= h(y;\twoth) = 
\theta_1\,(1-x-y^2) + \theta_2\,
(y - \epsilon);
\ee
\ni this region, call it $D$, is defined by:
\be
D \= \left\{\,(\twoth): h(y) \leq \half \ph \hbox{for all $y, -1 \leq y \leq 1$}\,\right\}.
\ee
We observe that $ D $ is the domain where $ c(\theta_1,\theta_2) $ is finite, related to the conditions for the validity of G\"artner-Ellis' theorem, not the domain of the logarithm argument.
The geometry of this region is a bit complicated. These are the tests for whether a point
lies in $D$:
\begin{quote} 

Test1: $h(1) \leq 1/2$;

Test2: $h(-1) \leq 1/2$;

Test3: if $y_c = \theta_2/2\theta_1$, the critical point of $h(y)$, 
	lies in [-1,1], then test whether $h(y_c) \leq 1/2$.

\end{quote}

Now see Figure 1. By tests one and two, $D$ lies between the lines A and B, 
and to the left of point P. Test three applies in between lines E and F; 
to the right of the axis, it is satisfied within
the oblique ellipse, curve C; to the left of the axis, $h$ is negative. Hence we obtain
a diamond-shaped compact region of the plane. 
\begin{figure}
	\centering\rotatebox{0}{\resizebox{3.8in}{3.8in}{\includegraphics{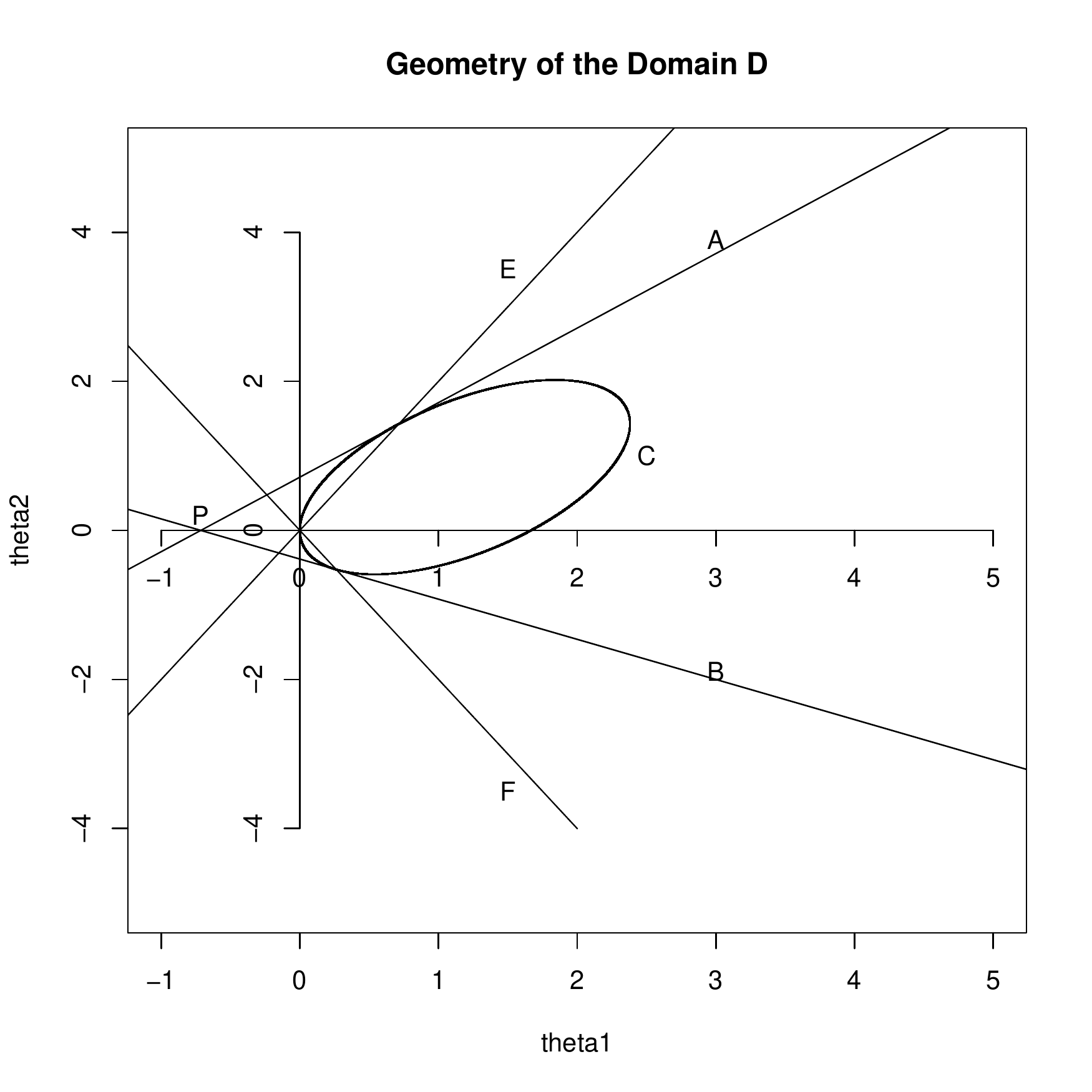}}}
	\caption{Regions used in describing the domain D where $c(\theta_1,\theta_2)$ is finite.}\label{fig:1}
\end{figure}
We note that $c(\twoth)$ is convex (by examination of its Hessian matrix; omitted) and ``steep",
meaning it's derivatives approach infinity at the boundaries of $D$.  We have an ellipse as in figure \ref{fig:1} if $ (1-x)-\epsilon^2>0 $, otherwise a hyperbola. We restrict the study to the ellipse case, since it is the relevant one: we are primary interested to small $ \epsilon>0 $ and from  the proof after \eqref{thenewratio}, one can see that the small $ x $ are the relevant ones for low temperatures. So given $ \epsilon $, we always restrict to the $ x $ small enough to satisfy $ (1-x)-\epsilon^2>0 $. 

The G-E procedure is to compute the dual function:
\be
f(z_1,z_2) \= \sup_{\twoth}\,\left[\, z_1\,\theta_1 + z_2\,\theta_2 - c(\twoth)\,\right];
\label{fdef}
\ee   
\ni from which you can compute the I-function as:
\be\label{defI}
I_2 \= I_2(x;\epsilon) \= \inf\,\left\{\,f(z_1,z_2): z_1 \leq 0; z_2 \geq 0\,\right\}.
\ee

By the convexity and steepness of $c$, the supremum in (\ref{fdef}) is attained at a critical
point for which
\bar
 \frac{\partial c}{\partial \theta_1} \= z_1;\,\,\,\,
 \frac{\partial c}{\partial \theta_2} \= z_2;
\ear
\ni so looking forward to a possible computer-search (for which we do not want to solve equations
but rather evaluate formulas), we define
\be
k(\twoth) \=  
 \frac{\partial c}{\partial \theta_1}\,\theta_1 \+  
 \frac{\partial c}{\partial \theta_2}\,\theta_2 - c(\twoth).
\ee

Let $G$ denote the constraint set in the $(\twoth)$-plane:

\be
G \= \left\{\,(\twoth): 
 \frac{\partial c}{\partial \theta_1} \leq 0; \, 
 \frac{\partial c}{\partial \theta_2} \geq 0\,\right\}.\label{constraint2}
 \ee 

\ni We can express the I-function as
\be\label{eq:I2k}
I_2 \= I_2(x;\epsilon) \= \inf\,\left\{\,k(\twoth): (\twoth) \in G\cap D;\,\right\}. 
\ee

For the proof we have to compare the I-function for the two-variable, two-inequality problem
with that of the one-variable, one inequality problem, given by
\be
I_1 \= I_1(x) \= \inf\,\left\{\,k(\theta_1,0): \theta_1 \in L;\, 
 \frac{\partial c}{\partial \theta_1}(\theta_1,0) \leq 0 \,\right\},\label{constraint1} 
\ee
\ni where $L$ is the part of the horizontal axis that lies in $D$. A first, critical
question to ask is whether necessarily $I_2 \leq I_1$ because the set over which we
compute the infimum for $I_2$ contains the set for which we compute the infimum for 
$I_1$. But this is not the case. We have:
\be
 \frac{\partial c}{\partial \theta_2} \= \int_{-1}^1\,dy\,\frac{y}{q(y)} \- \epsilon\, 
 \int_{-1}^1\,dy\,\frac{1}{q(y)},\label{par2}
\ee
\ni where $q(y) = 1 - 2\,h(y) \geq 0$ by the restriction that $(\twoth) \in D$. 
The first integral can be done easily by elementary calculus:
\def\intmo{{\int_{-1}^1}\,dy}
\bar
\no && \intmo\,\frac{y}{q(y)} \= \frac{1}{4\theta_1}\,\intmo\,\frac{4\theta_1\,y - 2\theta_2 + 2\theta_2}{2\theta_1\,y^2
- 2\theta_2 y + b} \=\\
\no && \frac{1}{4\theta_1}\,\log\left\{\,\frac{2\theta_1
- 2\theta_2 + b} 
{2\theta_1
+ 2\theta_2 + b}\,\right\} + \frac{\theta_2}{2\theta_1} \intmo\,\frac{1}{q(y)},\\
&&\label{easy}
\ear
Now note that, at $\theta_2 = 0$,
this integral is zero, and so is the first term in (\ref{par2}), while the second term there is
negative. Hence $\{\,\partial c/\partial \theta_2 \geq 0\,\}$ is disjoint from $L$.

We need a formula for the first partial of $c$:
\be
\frac{\partial c}{\partial \theta_1} \= \intmo\,\left\{\,\frac{1-x-y^2}{2\theta_1\,y^2 -
2\theta_2\,y + b}\,\right\}.
\ee
\ni By long division we can write:
\be
\frac{1-x-y^2}{2\theta_1\,y^2 -
2\theta_2\,y + b} \= A \+ \frac{B + C\,y}{q(y)},
\ee
\ni where
\be
A = -\frac{1}{2\,\theta_1}; \ph B = 1 - x - A\,b;\ph C = 2\,\theta_2\,A.
\ee
\ni Hence:
\bar
\no \frac{\partial c}{\partial \theta_1} &\=& \intmo\,A \+ B\,
\intmo\,\left\{\,\frac{1}{q(y)}\,\right\}
\+ C\,\intmo\,\left\{\,\frac{y}{q(y)}\,\right\}\\
\no &\=& 2\,A \+ \left(\,\frac{C\,\theta_2}{2\,\theta_1} + B\right)\,
\intmo\,\left\{\,\frac{1}{q(y)}\,\right\} \+
\frac{C}{4\,\theta_1}\log\left\{\,\frac{q(1)}{q(-1)}\,\right\}.\\
&&\label{parcone}
\ear
\ni where we have plugged in (\ref{easy}) for one integral.

Note that we have reduced computing these derivatives to computing one integral, of
$1/q(y)$ over the interval $[-1,1]$.
There are two cases, depending
on whether $q(y)$ is irreducible or factorizes. The discriminant is:
\bar
\no \hbox{disc.} \= 4\theta_2^2 - 8\theta_1\,b \= 4\theta_2^2 - 8\theta_1\,\left[\,1 - 2\,\theta_1(1-x) - \theta_2\,\epsilon\,\right].
\ear
\ni If disc. $ > 0$, $q$ factorizes as 
\be
q(y) = 2\theta_1\,(y-r_{+})\,(y - r_{-}),
\ee
\ni where the two roots are outside the interval $[-1,1]$. The integrals can then be performed by
partial fractions, yielding logarithmic terms:
\be
\intmo\,\left\{\,\frac{1}{q(y)}\,\right\} \= \frac{1}{2\,\theta_1\,(r_{-} - r_{+})}\,
\log\left[\,\frac{(1-r_{-})\,(1+r_{+})}{(1 - r_{+})\,(1+r_{-})}\,\right].\label{logcase}
\ee
If disc.$ <0$, in our case $q$ can be written:
\be
q(y) = A'(y-B')^2 + C',
\ee
\ni with $A' > 0$ and $C' > 0$, and one makes a trig substitution: 
\be
\tan(u) \= \sqrt{\frac{A'}{C'}}\,(y - B').
\ee
\ni The result contains inverse trig functions:
\be
\intmo\,\left\{\,\frac{1}{q(y)}\,\right\} \= \frac{1}{\sqrt{A'\,C'}}\,\left[\,
\arctan\left\{\,\sqrt{\frac{A'}{C'}}\,(1-B')\,\right\}  
\- \arctan\left\{\,\sqrt{\frac{A'}{C'}}\,(-1-B')\,\right\}\,\right].
\ee
To check these calculus formulas (and our computer implementation of them) we computed the integral directly (numerically) using the trapezoid rule and compared.
We also need a formula for $c$, which we can obtain by integrating-by-parts:
\bar
 \no c &\=& -\half\,\intmo\,\log\,q(y) \= -\half\,\intmo\,y'\,\log\,q(y)\\   
\no c &\=& -\half\,\log\left[\,q(1)\,q(-1)\,\right] \+ \half\,\intmo\,
\left\{\,\frac{y\,q'}{q}\,\right\}.\\
&&
\ear
 
\ni Here
\be
\half\,\intmo\,
\left\{\,\frac{y\,q'}{q}\,\right\} \= 
\intmo\,
\left\{\,\frac{2\,\theta_1\,y^2 - \theta_2\,y}{2\,\theta_1\,y^2 - 2\theta_2\,y + b}\,\right\},
\ee
\ni and applying long division again:
\be
\left\{\,\frac{2\,\theta_1\,y^2 - \theta_2\,y}{q}\,\right\} \= A \+ \frac{B\,y + C}{q},
\ee
\ni with
\be A = 1;\ph B = \theta_2; C = -b.
\ee
Thus we get
\be
\half\,\intmo\,
\left\{\,\frac{y\,q'}{q}\,\right\} \= 2 \+ \theta_2\,\intmo\,\left\{\,\frac{y}{q}\,\right\}
\- b\,\intmo\,\left\{\,\frac{1}{q}\,\right\}.
\ee
Combining with previous results yields:
\be
c \= 2 - \half\,\log\left[\,q(-1)\,q(1)\,\right] + \frac{\theta_2}{4\,\theta_1}\,\log\left[\,
\frac{q(1)}{q(-1)}\,\right] + \left\{\,\frac{\theta_2^2}{2\theta_1} - b\,\right\}
\intmo \left\{\,\frac{1}{q}\,\right\}.
\ee

For the function $k$ we can compute from its definition, but also some simplifications accrue:
\bar
\no k(\twoth) &\=&  
 \frac{\partial c}{\partial \theta_1}\,\theta_1 \+  
 \frac{\partial c}{\partial \theta_2}\,\theta_2 - c(\twoth)\\
\no &\=& \intmo\,\left\{\,\frac{\theta_1(1-x-y^2)}{q(y)} \+ \frac{\theta_2(y-\epsilon)}{q(y)}\,
\right\} \- c(\twoth)\\
\no &\=& \intmo\,\left\{\,\frac{h(y)}{1-2h(y)}\,\right\} - c(\twoth)\\
\no &\=& - \half\,\intmo\,\left\{\,\frac{-2h(y) + 1 -1}{1-2h(y)}\,\right\} - c(\twoth)\\
\no &\=& -1 \+ \half\,\intmo\,\left\{\,\frac{1}{q(y)}\,\right\} \+ \half\,\intmo\,
\log(q(y)).\\\label{kformula}
&&
\ear
\ni For the theorem we have to evaluate the asymptotics of the ratio: 
\be
\frac{ \exp\{\,-N\,\beta\,\}\,|B| + N\,\beta\,\int_0^1\,dx \exp\{\,- N\,g_2(x)\,\}}
{ \exp\{\,-N\,\beta\,\} + N\,\beta\,\int_0^1\,dx \exp\{\,- N\,g_1(x)\,\}},\label{theratio}
\ee
\ni where $B$ is the set appearing in (\ref{setB}) and
\be
g_{1,2}(x) \= \beta\,x + I_{1,2}(x).
\ee

We next list some properties of the $I$-functions that will imply the theorem's conclusion:
\begin{quote}
\centerline{Assumptions on $I_2(x)$ and $I_1(x)$}
\bar
\no &\hbox{(a)}& \ph I_2(x) > I_1(x) \ph\hbox{for}\ph x \in (0,1];\\
\no &\hbox{(b)}& \ph I_{1,2}\ph\hbox{are differentiable, monotone decreasing,
 and convex on (0,1]};\\
\no &\hbox{(c)}& \ph\lim_{x \to 0}\, I_{1,2}(x) = + \infty;\\ 
\no &\hbox{(d)}& \ph I_2(1) > 0;\\ 
\no &\hbox{(e)}& \ph I_1(x) = 0 \ph \hbox{for}\ph x \in [2/3,1].\\
&&
\ear
\end{quote}
The rate functionals are defined as 
\be
I_{1,2}(x) \= - \lim_{N \to \infty}\,1/N\,\log\,P_{1,2;N}(x).
\ee
\ni
Since they are computed from  independent Gaussian variables, even if not identically distributed \cite{PR},  the corrections to the large deviations are $  O\left(N^{-{1/2}}\right) $: 
\begin{equation}\label{eq:LD correction}
P_{1,2;N}(x) = \exp\left[-N\left(I_{1,2}(x)+O\left(\frac{\log N}{N}h_{1,2}(x)\right) \right)\right].
\end{equation}
Therefore  the approximation $ P_{1,2;N}(x) \approx \exp\left[-NI_{1,2}(x)\right]$ is  very good for large $ N $. 
If $ I_2(x)\neq I_1(x) $ then it has to be $ I_2(x)>I_1(x) $.  The study of their large deviation corrections should prove it.

For  (a)  we know from the inequality: $P_{2;N}(x) < P_{1;N}(x)$  (for fixed $ N $) and the \Gartner-Ellis Theorem $ I_2(x)\geq I_1(x) $,
and  we showed that these functions result from infimums of the same function over
different sets; therefore, it is implausible that they are exactly equal 
for all $x$ smaller than a given\footnote{Because of the property b) and $ I_2(x)\geq I_1(x) $, if there exists a $ x^* $ such that $ I_1(x)=I_2(x) $, then they have to coincide for all $ x $ smaller than $ x^* $.} $ x^* $. (We supply some evidence that it is not so, from numerical approximations, 
see the Computational Appendix \ref{app:numerics} and Figure Three.). 
From the approximation $ P_{1,2;N}(x) \approx \exp\left[-NI_{1,2(x)}\right]$ and $ P_{i;N}(x')> P_{i;N}(x) $ for $ x'>x $ one  expects strict decreasing monotonicity. This is indeed what we observed from the computer computations. Differentiability and strict decreasing monotonicity can be deduced from  the  Envelope Theorem\footnote{A theorem very  popular in Mathematical Economics, mainly related to the optimization  of the value function. }  in \cite{Carter} at pag. 605 applied to    \eqref{fdef} and \eqref{defI}.   $ f(z_1,z_2) $ is  strictly convex and also the Lagrangian  $ f(z_1, z_2)- \lambda\cdot z  $ to optimize if the optimizer is on the boundaries. Applying the Envelope Theorem,  on the right hand side of \eqref{eq:envelope} the derivative coincides with that of $ f $  because  the constraint functions doesn't have dependence on   $ x $.  From $ \underset{z}{\inf}\,\, f(z_1,z_2)$,  calling $ \theta^*(z) $ the solution of \eqref{fdef} and $ z^* $ the one of \eqref{defI}, we have 
\begin{eqnarray}
\frac{d I_i}{dx}=\frac{\partial f(\theta^*(z^*(x),x),z^*(x),x)}{\partial x}\label{eq:envelope}.
\end{eqnarray}
The smoothness of $ \theta^* $ and $ z^* $ comes from the implicit function theorem.  
The right hand side gives $ \theta^*_1\int_{-1}^{1} \frac{dy}{q} $, since $ q>0 $ we have that it is negative if and only if $ \theta_1<0 $. We know that $ I_i(x) $ is decreasing by definition so it can be only $ \theta_1 <0 $ or $ \theta_1=0 $, for $ I_1(x) $ this second case gives $ I_1(x)=0 $, which it can be only for $ x\geq 2/3 $ by definition, therefore $ \theta_1^*<0 $ when $ I_1(x) $ is not trivial. For $ I_2(x) $ the value $ \theta_1^* $ stays in disjoint set  from the axis $ \theta_1=0 $, therefore it can be only $ \theta^*_1<0 $ since $ I_2(x) $ is decreasing by definition. Second order differentiability needs to develop second order Envelope Theorems, of course it will not be $ \frac{d^2 I_i}{dx^2}=\frac{\partial^2 f}{\partial x^2} $. Differentiability of the rate functionals can also be deduced from differentiability of the probability functions $ P_{i;N}(x) $, in our case one can apply corollary 4.1 of \cite{PV1} and corollary 32 of \cite{PV2} because of the quadratic structure  of the $ \chi_n $'s and other conditions that are verified. These results should be  extended to  second order differentiability without problem \cite{vA}. 
We can not deduce analytically the convexity property ( varying $ x $ we are restricting in a continuous way the integration on regular set of a Gaussian integral) and  we observe this property from the numerical computations of the exact formulas of the rate functionals, see appendix \ref{app:numerics},  within the limits of our computer analysis, that is below $ x=0.1 $. Here we can not obtain good numerical estimates since the sampling region shrinks to an infinitesimal volume and  we have a vertical asymptote, as proved later for (c). In this vertical asymptote part convexity is natural.  Anyway from the proof after \eqref{thenewratio}, if there is some flex from some unexpected reason, the computations can be applied for smaller $ x $ (i.e. smaller temperature) where the functionals are convex. From  \eqref{setB} we have (d) since the first event for $ x=1 $ is the sure event, while the second  never contains the average for any $ \varepsilon>0 $ (see also the Computational Appendix \ref{app:numerics} for a numerical computation).

Also, (e) follows from a computation
in a previous section, see (\ref{gnlimit}), and the remark that LD results from probabilities
of sets that do not contain the mean; otherwise, by the Central Limit Theorem
such probabilities go to one. 

\def\intmoneone{{\int_{-1}^1\,dy}}

That leaves (c). It would follow immediately from the observation that 
$P_{1,2}(0) = 0$ (since $g_n \leq 1$)
if we can assume that two limits can be interchanged. As that is not obvious,
we supply a proof using previously-obtained formulas. 
Since certainly $I_2(x) \geq I_1(x)$ it suffices to prove (c) for the latter.
So we set $\theta_2=0$ in the following. From (\ref{parcone}) we note that
\be
\frac{\partial c}{\partial \theta_1} \= \frac{1}{\theta_1}\,\left\{\,
-1 + \half\,\int_{-1}^1\,\frac{1}{q(y)}\,\right\}.\label{cset}
\ee
Let's define for a given $x$:
\be
H \= H(\theta_1) \= 
-1 + \half\,\intmoneone\,\frac{1}{q(y)}.
\ee
\ni We note $H(0) = 0$, since $q = 1$ there, and
\bar
\no \frac{\partial H}{\partial \theta_1}(0) &\=& - \half\,\intmoneone\,
\left\{2\,y^2 - 2(1-x)\right\}\\
\no &\=& \frac{4}{3} - 2\,x;\\
&&
\ear
\ni so, if $x < 2/3$, $\partial H/\partial \theta_1 > 0$. 
Also:

\be
\frac{\partial^2 H}{\partial \theta_1^2} \=  \intmoneone\,\frac{\left\{
2y^2 - 2(1-x)\,\right\}^2}{q(y)^3} \geq 0,
\ee

\ni since by definition of the domain D, $q(y) \geq 0$ for all $y$ in [-1,1].
Hence, $H$ is convex.

Now consider the
left-most point of the domain D on the horizontal axis, labeled `P' in Figure One,
which has $\theta_1$ coordinate $-1/2x$.
By definition, as $\theta_1 \to \hbox{P}$, $q(y) \to 0$ for some $y \in [-1,1]$ 
(here, at the endpoints $y = \pm1$), 
so $H \to \infty$. Thus
$H(\theta_1)$ must have a second zero on the line, to the right of $P$; let's label it $Q$. 

The constraint set is $H \geq 0$, so from (\ref{cset}) we conclude that it lies
entirely to the left of $Q$ (and of course to the right of $P$). 

Some computer work indicated that $Q$ is very close to $P$ for small $x$, which explains
why we could not locate the constraint set by sampling for $x < 0.1$.
Hence, suppose that $Q(x) = P(x) + \delta(x)$ with $\delta(x)$ bounded above  
 (or even, as we shall see, tends to zero) as $x \to 0$.
From (\ref{kformula})
we have 
\be
k(\theta_1,0) \= H + \half\,\intmoneone\,\log\{q(y)\}.
\ee
Note that, substituting $-1/2x + \eta$ for $\theta_1$,
\be
q(y) \= \frac{1 - y^2}{x} + 2\,\eta\,y^2 - 2\,\eta\,(1-x).
\ee
\ni So in the constraint set with the above assumption $\eta \leq \delta(x)$,  
so the second term in $k$ goes to infinity as $x \to 0$.
Hence, since the first term is nonnegative by the constraint,
\be
\inf\,\{\,k(\theta_1,0):\, \theta_1\, \in\, D\,\} \to \infty,
\ee
\ni as $x \to 0$, which yields (c).

In order to prove that $Q(x) \to P(x)$ we apply the exact formula for the integral
appearing in $H(\theta_1)$, which, with $\theta_2 = 0$, lies in the factorizable case,
see (\ref{logcase}). Letting
\be
s = \sqrt{\,\frac{1}{-2\theta_1} + 1 - x\,},
\ee
\ni we find
\be
H \= \left(\frac{s^2 + x -1}{4\,s}\right)\,\log\left\{\frac{(1+s)^2}{(1-s)^2}\right\} - 1.
\ee

Next, let $t = s -1$, so that $t = 0$ corresponds to $\theta_1 = -1/2x$; i.e.,
$P(x) = Q(x)$. Rewriting the equation for the root $H=0$ gives:
\be
\log(t) \= \log(2+t) - \frac{2(t+1)}{t^2 + 2y +x};
\ee
\ni so exponentiating both sides:
\be
t = (2+t)\,\exp\left\{\,
\frac{-2(t+1)}{t^2 + 2t +x}\,\right\}.\label{teq}
\ee
Note that, if $x$ is small, the right-hand side of (\ref{teq}) is very small at $t=0$
and remains so up to small values of $t$, while the left-hand side follows $t$.
For instance, if $x = .01$, up to $t = .01$ the RHS is no more than about $\exp\{
-66\}$ which is infinitesimal while the LHS reaches $.01$. 
Thus, as $x \to 0$, the solution of (\ref{teq}) goes to zero, 
and hence $Q(x) \to P(x)$ (in fact exponentially fast),
which implies (c).

Granted these assumptions, we now proceed to the proof of Theorem Two.
Property (e) implies that we can replace the denominator in (\ref{theratio})
as follows:

\be
\frac{ \exp\{\,-N\,\beta\,\}\,|B| + N\,\beta\,\int_0^1\,dx \exp\{\,- N\,g_2(x)\,\}}
{ \exp\{\,-(2/3)\,N\,\beta\,\} + N\,\beta\,\int_0^{2/3}\,
dx \exp\{\,- N\,g_1(x)\,\}},\label{thenewratio}
\ee

Next, consider whether the $g$-functions have critical points in the intervals (0,1) or (0,2/3).
Using a hat to denote those points, we are asking whether solutions exist for:
\be
\beta + I'_1(\hat{x}_1) = 0;\ph\ph 
\beta + I'_2(\hat{x}_2) = 0,\ph\ph 
\ee
First, suppose both exist. If so, since 
$g$ is convex, the Laplace approximation applies yielding:
\be
\int\,\exp\{\, - N\,g(x)\,\} \approx \frac{\sqrt{2\,\pi}\,\exp\{ - N\,\hat{g}\,\}}
{\sqrt{N\,\hat{g}''}},
\ee
\ni where a hat means evaluate at the critical point.
We are now prepared to argue that the limit of the ratio in (\ref{thenewratio}) is always zero.
There are four cases, defined by whether the ``winner" (largest term asymptotically) in numerator
and denominator is the first or second term:
\begin{quote}

\centerline{{\bf Case}: numerator, second term; denominator, second term.} 
As the $\hat{g}$'s are minimums, $\hat{g}_2 > \hat{g}_1$ and the ratio tends to zero.
\vskip0.1in
  
\centerline{{\bf Case}: numerator,  first term; denominator, second term.}  
 Then $\hat{g}_1 < 2/3\,\beta$ and we are looking essentially at the limit of
\be
\frac{\exp\{ - N\,\beta\,\}\,|B|}{\exp\{ - N\,\hat{g}_1\,\}}
\ee
\ni which clearly goes to zero.
\vskip0.1in

\centerline{{\bf Case}: numerator, second term; denominator, first term.}  
  Now we are looking at
\be
\frac{\exp\{ - N\,\hat{g}_2\,\}\,}{\exp\{ - (2/3)\,N\,\beta \, \}}
\ee
\ni which goes to zero if $\hat{g}_2 > 2/3\,\beta$. The reverse is impossible,
as the ratio would tend to infinity while we know it is bounded by one from the original
definition.
\vskip0.1in
 
\centerline{{\bf Case}: numerator, first term; denominator, first term.} 
 The ratio tends to zero.
\vskip0.1in
\end{quote}

Since we are interested in low temperatures (that is small $ x $) the presented cases are the relevant ones. But we can still ask: do these critical points exist? With our assumptions we know that 

\be
\lim_{x \to 0}\,I'_{1,2}(x) = - \infty.
\ee

\ni If in addition we knew that

\be
\lim_{x \to 1}\,I'_2(x) = 0 \ph \hbox{and} \ph 
\lim_{x \to 2/3}\,I'_1(x) = 0
\ee

\ni we can conclude the c.p.'s exist. If not, for sufficiently small $\beta$
 a c.p. might not exist. For instance, let us postulate that $I_2'(1) = - c < 0$ or
 $I_1'(2/3) = - c < 0$.
Then in either case  
\be
g'(x) \leq \beta - c < 0,
\ee
\ni for all $x$ in the relevant interval. In this case $g$ is monotonic and we can make a change of variable: $u = g(x)$ to obtain
\bar
\no  \beta\,N\,\int_0^1\,dx\,\exp\{ - N\,g(x)\,\} &\=&
 \beta\,N\,\int_{g(1)}^{g(0)}\,du\,\left[\,\frac{1}{-g'\circ g^{-1}(u)}\,\right]
\,\exp\{\,- Nu\,\}\\
\no &\leq&  \beta\,N\,\int_{g(1)}^{g(0)}\,du\,\left[\,\frac{1}{c - \beta}\,\right]
\,\exp\{\,- Nu\,\}\\
&&
\ear
So for the numerator the integral is:
\be
\leq \left[\,\frac{1}{c - \beta}\,\right]\,\exp\{ - N\,[\beta + I_2(1)]\,\}
\ee
\ni (where we have used assumption (c) to drop a term) and so, if $I_2(1) > 0$, the first term in the numerator wins. In the denominator, the order is 
at most $\exp\{ - 2/3\,\beta\,N\,\}$,
so the numerator tends to zero faster whatever term predominates in the denominator. 

CAQED (``Computer-Assisted QED").

\section{Math Appendix: SCW with wavefunction energy}\label{app:SCWwithWFE}

\def\ooN{{\frac{1}{N}}}
\def\sdel{{\sqrt{\delta}}}
\def\mhalf{{-\frac{1}{2}}}
\def\Amax{{A_{\hbox{max}}}}
\def\Amin{{A_{\hbox{min}}}}

We prove here that, assuming the hypotheses of Theorem Three, if
\be\label{eq:critemp}
\beta > \frac{\underline{p}^*(\epsilon)}{(\omega - 1)\epsilon},
\ee
\ni where $\underline{p}^*(\epsilon)$ is a certain positive function specified in the proof below, then the conclusion of the theorem follows.

We will apply the Concentration Lemma.
We can assume by adding a term as usual to render $f$ positive:

\be
f \= N\,\beta\,\left\{\, 1 - m^2(\phi) + (\omega - 1)\,D(\phi)\,\right\}.\label{fform}
\ee

We next define the quantities involved in the Lemma. For the sets $U$ and $V$ we take,
given two positive numbers $\epsilon$ and $\eta$ ($\eta$ may depend on $N$),
\bar
 U \= \left\{\,\phi:\ph m^2(\phi)\,\geq\, \epsilon\,\right\};\,\,\,\, V \= \left\{\,f \leq \,\eta\,\right\}.
\ear
\ni Further, define:
\be
\alpha \= \beta\,N\,\left(\,1 - \epsilon\,\right).
\ee
\ni We next check the hypotheses of the Concentration Lemma.
A quick calculation gives the equivalent
description of set $V$:
\bar
\no V &\=& \left\{\,\phi:\ph\ph \omega\,m^2(\phi) \geq (\omega - 1)\,
\sum\,|\phi_n|^2\,g_n^2\,
+\,(1-\eta/\beta N)\,\right\}\\
\no  &\=& \left\{\,\phi:\ph\ph m^2(\phi) \geq
\sum\,|\phi_n|^2\,a_n
\,\right\},\\
&&
\ear
\ni where 
\be
g_n = 1 - \frac{2\,n}{N}
\ee
\ni and we have defined
\bar
\no a_n \= \frac{\omega - 1}{\omega}\,g_n^2 \+ \frac{1-\eta'/\beta}{\omega}
\no \= r\,g_n^2 \+ \delta.
\ear
Since $\omega > 1$, assuming $\eta = \eta'\,N$ and $1 - \eta'/\beta > \omega\,\epsilon$,
the condition defining $V$ implies $V \subset U$.
Since $\omega > 1$, it follows from (\ref{fform}) that $f \geq \alpha$ on $U^c$.

For the lower bound on $|V|$, we translate to a model with i.i.d. Gaussians, call them
$\chi_n$, replacing:

\be
\phi_n \longrightarrow\, \frac{\chi_n}{\sqrt{\sum |\chi_n|^2}},
\ee

\ni and the definition of $V$ becomes:

\be
V \= \left\{\,\left(\,\sumnN\,|\chi_n|^2\,g_n\,\right)^2 \geq \left[\,\sumnN\,a_n\,|\chi_n|^2\,
\,\right]\,\sumnN\,|\chi_n|^2\,\right\}.
\ee

\ni (Since $\psi_n$ has both a real and an imaginary part, 
we should double the lengths of these sums, but clearly this has no impact on the final result.)
We have to find an exponential lower bound on $P[V]$.
Since $a_n \geq \delta$, we can get a lower bound by writing:
\bar
\no && P\left[\,V\,\right] \geq  P\left[\,\left(\,\ooN\sumnN\,|\chi_n|^2\,g_n\,\right)^2 \geq 
\left\{\,\ooN\sumnN\,a_n\,|\chi_n|^2\,
\,\right\}\,\ooN\sumnN\,\frac{a_n}{\delta}\,|\chi_n|^2\,\right] \=\\
\no && P\left[\,\sqrt{\delta}\,\ooN\sumnN\,|\chi_n|^2\,g_n\, \geq \ooN\sumnN\,a_n\,|\chi_n|^2\,
\right] \+ P\left[\,\sqrt{\delta}\,\ooN\sumnN\,|\chi_n|^2\,g_n\, \leq - \ooN\sumnN\,a_n\,|\chi_n|^2\,
\right].\\
&&
\ear 
We'll work on the first probability on the last line. (The second one is really the same since,
multiplying the inequality through by -1, replacing $g_n$ by $-g_n$ just reverses
the order of its values.) 
According to \Gartner-Ellis, we have to compute:
\be
p(\theta) \= \lim_{N\to\infty}\,\ooN\,\log\,{\cal{E}}\exp\{\,\theta\,\sum\,\chi_n^2\,b_n\,\},
\ee
\ni where $\cal{E}$ denotes expectation and 
\bar
 b_n \= \sdel\,g_n - a_n \= \sdel\,\left(\,1 - \frac{2n}{N}\,\right) - r
\,\left(\,1 - \frac{2n}{N}\,\right)^2 - \delta.
\ear

Note that $b_n$ can take negative and positive values; hence, 
$\theta$ must be restricted to ensure that the integral is finite. 
From the standard Gaussian integral we get:
\be
p(\theta) \= \mhalf\, \lim_{N\to\infty}\,\sum\,\log\left( 1 - 2\,\theta\,b_n\,\right),
\ee
\ni which, provided the integrand is bounded, we recognize as the Reimann sum 
converging to:
\be
p(\theta) \= - \frac{1}{2}\,\int_0^1\,du\,\log\left( 1 - 2\,\theta\,A(1 - 2u)\,\right),
\ee
\ni where
\be
A(x) \= \sdel\,x - 
r\,x^2 \,\- \delta.
\ee
By a change of variable this equals:
\be
p(\theta) \= - \frac{1}{4}\,\int_{-1}^{1}\,dx\,\log\left( 1 - 2\,\theta\,A(x)\,\right).
\ee
The LD approach requires us to compute:
\bar
 p^*(y) \= \sup_{\theta}\,\left\{\,\theta\,y - p(\theta)\,\right\}; \underline{p}^* \= \inf_{y > 0}\,p^*(y);
\ear
and then the asympotic lower bound is $\exp(-\underline{p}^*\,N\,)$, \cite{ellis}. 

By the definition of the domain where $ p(\theta)<+\infty $ in G\"artner-Ellis' theorem, the supremum over $\theta$ in the definition of $p^*(y)$ can be limited to:
\be
  \frac{1}{2\,\Amin} \leq \theta \leq \frac{1}{2\,\Amax},
\ee
\ni where $\Amin$ is negative but we don't need to know it, while by a simple computation:
\be
\Amax \= \delta\,\left\{\,\frac{4 - 3\omega}{4(\omega -1)}\right\}.
\ee
\ni (This is the max on the whole line).
We must have positive values of $A(x)$ somewhere in the interval [-1,1], for otherwise 
$\lim\,p(\theta) = - \infty$ as $\theta \to \infty$, so $p^*(x) = +\infty$ for $x \geq 0$.
This requires $\Amax >0$ and that the lower root of $A(x) = 0$ lies in the interval [0,1] (since
$A(0) = - \delta < 0$ and $A'(0) > 0$).
The root is easily computed to be:
\be
x_{-} \= \frac{\sdel\,\left(\,1 - \sqrt{1 - 4\,r}\,\right)}{2\,r},
\ee
\ni Since $r < 1/4$,this root is real; 
letting $u = \sqrt{1 - 4\,r}$ the condition $x_{-} < 1$ becomes 
\be
\delta < \frac{1}{4}\,\left(\, 1 + u\,\right)^2,
\ee
\ni Since we will eventually identify $\delta$ with $\epsilon$,
the above inequality is identical with (\ref{epineq}). Since $0\leq u \leq 1$, the infimum of the
right side is 1/4.

The problem of computing $p^*(x)$ is then well defined. 
We note that, if the point at which $A(x) = \Amax$ lies in the unit interval, 
then as $\theta \to 1/2\Amax$, the integral looks as if it might diverge. 
However, it remains finite, because logarithmic singularities are integrable.
E.g., 
\bar
\no \int_0^1 dx\,\log(x) &\=& \lim_{\epsilon \to 0}\,\int_{\epsilon}^1\,dx\,\log(x)\= \lim_{\epsilon \to 0}\,\,\left[\,x\log(x) - x\,\right]_{\epsilon}^1 \= -1,
\ear
\ni not $- \infty$. However, the derivative $p'(\theta)$ will go to infinity at the boundaries
(Ellis calls such a function ``steep" and it is an assumption of his theorem).

Assuming that $0 < \underline{p}^* < \infty$, it will suffice for Theorem Three to know that
for some $\eta' > 0$:
\bar
 \underline{p}^* + \eta' \leq \beta\,(\,1 - \epsilon\,);\,\,\,\, \eta' \leq \beta\,(\, 1 - \omega\,\epsilon\,); \label{ineq}
\ear
\ni We can define $\eta'$ to saturate the second inequality above 
(note that $\omega\,\epsilon < 1$), which also yields
$\delta = \epsilon$. 
The conclusion of the theorem follows. QED

Can we compute $p(\theta)$ and $\underline{p}^*(y)$? In fact, the integral defining $p(\theta)$
is elementary and can be computed as follows, First, integrate by parts:
\bar
\no p(\theta) &\=& - \frac{1}{4}\,\int_{-1}^1\, dx\,\log\left\{\,1
- 2\,\theta\,A(x)\,\right\}\\
\no  &\=& - \frac{1}{4}\,\int_{-1}^1\, dx\,x'\,\log\left\{\,1
- 2\,\theta\,A(x)\,\right\}\\
\no &&  - \frac{1}{4}\,\left[\,x\,\log\left\{\,1
- 2\,\theta\,A(x)\,\right\}\right]_{-1}^1 \-  \frac{\theta}{2}\,\int_{-1}^1\, dx\,\frac{x\,A'(x)}{\,1
- 2\,\theta\,A(x)\,}.\\
&&
\ear
\ni There are two cases, depending on whether the denominator
in the integrand factorizes or not. For $\theta$ positive, it does not; hence the denominator
is an irreducible quadratic. For some negative $\theta$ it does factorize. We can proceed
by elementary linear and trig substitutions, yielding either log(linear), log(quadratic),
or arctan(quadratic) terms. 
Clearly, given such formulas with nonalgebraic functions, computing the supremum cannot
be expected in closed form, so we would again resort to the computer, but do not report
detailed results here. 

\section{Computational Appendix}\label{app:numerics}

After implementing the functions appearing in Math Appendix \ref{app:SCWnoWFE}
in a program, we employed a sampling scheme to
locate the constraint set. The idea is to choose at random
a ray emanating from point `P' of Figure One, then a point on the ray staying
within the domain $D$. (We also tried a ray emanating from the origin, which gave similar results.)

By experiment, we discovered that for $x$ near zero
the region G shrank to nearly a line close to line A in Figure One,
while the constraint set for generating $I_1$ lay almost at the left endpoint, 'P'. 
Hence we needed sampling schemes that could be biased to prefer points 
near the endpoints of a given interval [a,b] of the real axis.
Such schemes are given by: to bias near b, choose a point `$s$' by the scheme
\bar
\no s &\=& a + \log(z\eta\,u + 1)/\eta\\
\no z &\=& (\,\exp\{\eta\,(b-a)\,\} -1\,)/\eta.\\
&&
\ear
and to bias near a:
\be
s = a - \log( 1 - z\,\eta\,u\exp\{-\eta\,(b-a)\}\,)/\eta,
\ee
\def\bias{{\hbox{biased-sample}}}
\ni with the same expression for $z$. Plugging in a uniform random variable (produced by the
system RNG) for $u$ yields the sampling scheme for $s \in [a,b]$. 
With $\eta = 0$ the sampling is uniform on the
interval; large positive $\eta$ yields bias. 
We will write: $s = \bias(a,b)$ for the random sample.

Our sampling scheme in the region $D$ was the following:

\begin{quote}
(1) Choose $\alpha$: $\alpha = \bias(\,-x/(1+\epsilon),x/(1-\epsilon)\,)$;

(2) Choose $\theta_1$: $\theta_1 = \bias(\,-1/(2x), 5/(1-x)\,)$;

(3) Let $\theta_2 = \alpha\,[\theta_1 + 1/(2x)]$.
\end{quote}

If the selected $(\twoth)$ pass all the tests to lie in $D$, 
accept the values; otherwise, reject them.
To locate the constraint set, we sampled many pairs $(\twoth)$ as above and evaluated the
two partial derivatives of $c$, keeping and plotting the points that had the required
signs. The results---see Figure 2; 
parameters in the figure were $x = 0.7$ and $\epsilon = 0.3$---show 
that $G$ lies in the upper half of the plane,
and is disjoint from the horizontal axis.
Figure 3 shows curves of the I-functions obtained by sampling for a few values
of $x$ (0.1 to 0.7, in increments of 0.1). We used 10 billion samples at each $x$-value. 
Unfortunately,
we were unable to estimate the I-functions for $x < 0.1$ because few or no sampled points
fell in the constraint set (even with various choices of bias), for reason indicated earlier.
\begin{figure}[H]
	\centering\rotatebox{0}{\resizebox{3.5in}{3.5in}{\includegraphics{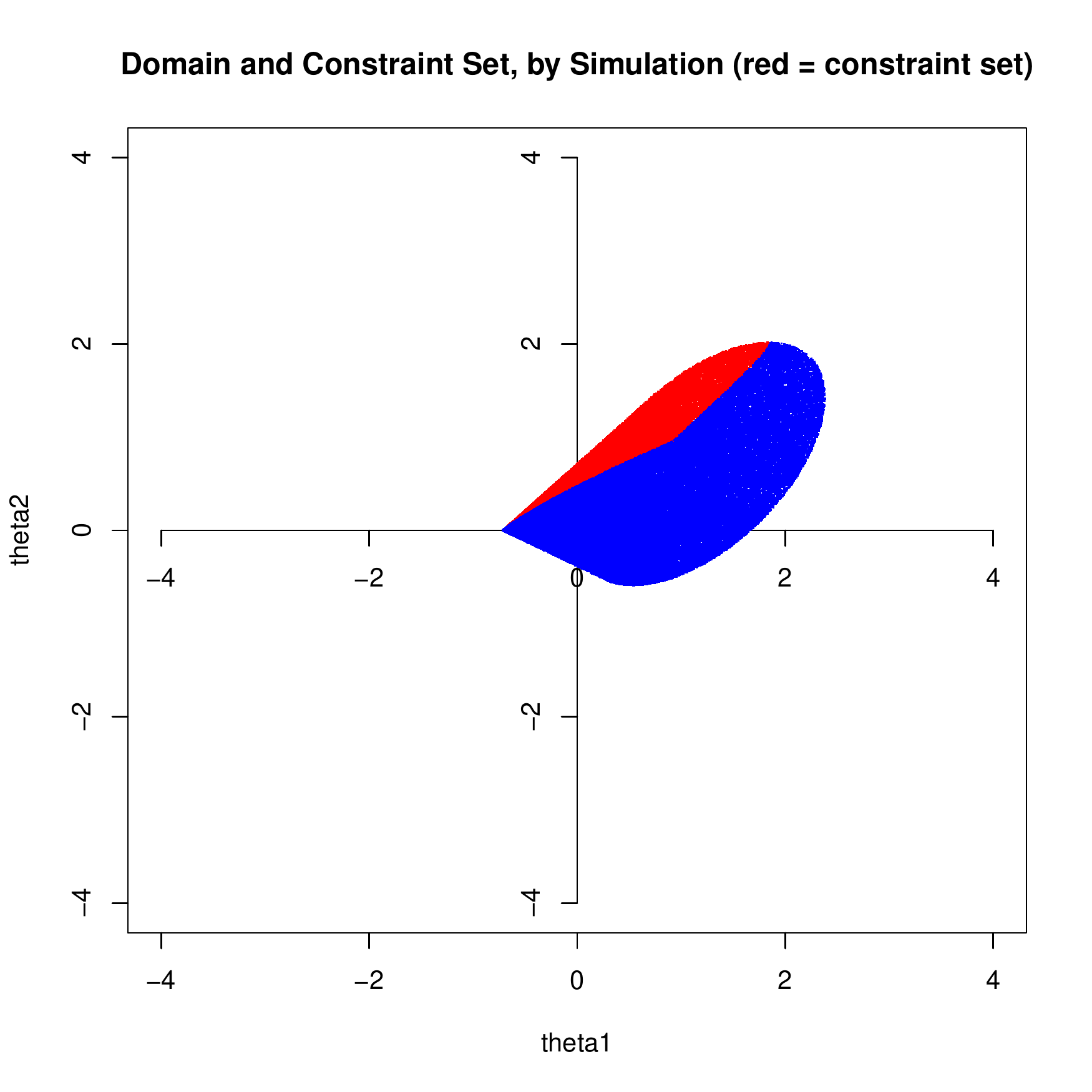}}}
	\caption{The Domain and Constraint Set located by sampling.}
\end{figure}
\begin{figure}[H]
\centering\rotatebox{0}{\resizebox{3.5in}{3.5in}{\includegraphics{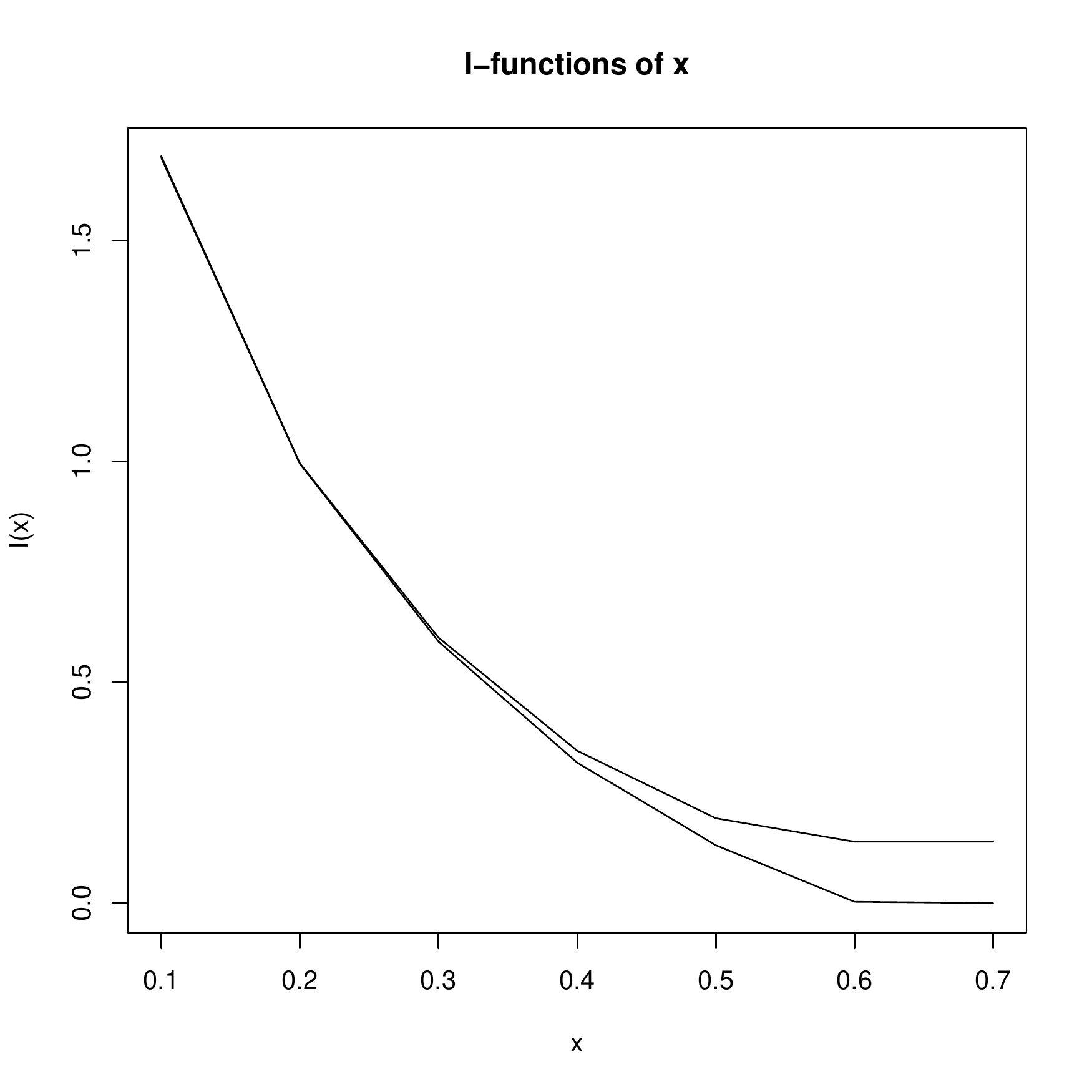}}}
\caption{I-functions obtained by sampling. The upper one is the I-function appearing in
 the numerator.}
\end{figure}

\subsection*{GitHub repository}

The code and header files used for the computational part is available in the GitHub repository: \textcolor{blue}{\href{https://github.com/leodecarlo/Computing-Large-Deviation-Functionals-of-not-identically-distributed-independent-random-variables}{Computing-Large-Deviation-Functionals-of-not-identically-distributed-independent-random-variables  }}

\subsection*{Acknowledgements}

Leonardo De Carlo thanks Federico De Iure and the ward  Chirurgia Vertebrale  Ospedale Maggiore Bologna to have  back a normal life, the DEF of LUISS Guido Carli di Roma and prof. Fausto Gozzi for the allowed time to finish the writing of the present project and Laura with her cats for the after-launch-time  during the solitary Winter in Tirrenia(PI).

\end{document}